\documentclass[12pt]{article}

\usepackage{graphicx}
\usepackage{amssymb}
\usepackage{epstopdf}
\usepackage{url}
\usepackage{mathpazo}

   \usepackage{rotating}
\newcommand{\bit}{\begin{itemize}} \newcommand{\eit}{\end{itemize}}
\newcommand{\im}{\item}

\newcommand{\ignore}[1]{}
\newcommand{\be}{\begin{equation}} \newcommand{\ee}{\end{equation}}
\newcommand{\ba}{\begin{eqnarray}} \newcommand{\ea}{\end{eqnarray}}
\newcommand{\nn}{\nonumber} \renewcommand{\bf}{\textbf}
\newcommand{\ra}{\rightarrow}

\renewcommand{\a}{\alpha}

 \renewcommand{\i}{\mathrm{i}}
\newcommand{\p}{\partial}  
  \newcommand{\NN}{\vec \nabla}

\def\slasha#1{\setbox0=\hbox{$#1$}#1\hskip-\wd0\hbox 
to\wd0{\hss\sl/\/\hss}}
\def\slashb#1{\setbox0=\hbox{$#1$}#1\hskip-\wd0\dimen0=5pt\advance
        \dimen0 by-\ht0\advance\dimen0 by\dp0\lower0.5\dimen0\hbox
          to\wd0{\hss\sl/\/\hss}}

\input{epsf}

\date{}
\begin{document}

\title{Quantum Theory  \\ without Planck's Constant}

\author{John P. Ralston \\   
{ \it Department of Physics \& Astronomy } \\ {\it The University of Kansas, Lawrence KS 66045  }}
\maketitle 

\begin{abstract} \small Planck's constant was introduced as a fundamental scale in the early history of quantum mechanics. We find a modern approach where Planck's constant is absent: it is unobservable except as a constant of human convention. Despite long reference to experiment, review shows that Planck's constant cannot be obtained from the data of Ryberg, Davisson and Germer, Compton, or that used by Planck himself. In the new approach Planck's constant is tied to macroscopic conventions of Newtonian origin, which are dispensable. The precision of other fundamental constants is substantially improved by eliminating Planck's constant. The electron mass is determined about 67 times more precisely, and the unit of electric charge determined 139 times more precisely. Improvement in the experimental value of the fine structure constant allows new types of experiment to be compared towards finding ``new physics.''  The long-standing goal of eliminating reliance on the artifact known as the International Prototype Kilogram can be accomplished to assist progress in fundamental physics.  

\end{abstract}

\section{The Evolution of Physical Constants }

More than a century after Planck's work, why are we still using Planck's constant? The answer seems self-evident. Planck's discovery led to quantum mechanics, which explains experimental data. Yet there exists a picture of quantum theory where Planck's constant is spurious. It cannot by found by fitting quantum mechanics to data. 

I propose that Planck's constant originates in human conventions. The conventions come from an era assuming Newtonian theory was fundamental. The relevance of the macroscopic theory has faded over time, and it no longer constitutes first principles. The quantum theory that replaced it turns out to have an unrecognized degree of symmetry. Assuming quantum mechanics is primary and fundamental, the symmetries reveal certain Newtonian conventions that fundamental physics can do without. 

It is often thought that fundamental constants have an absolute experimental character, which must override theory. The electron mass $m_{e}$ is an example. By a certain trick it has been planted as pre-existing and well-defined, with one invariant value in several different theories. In our approach no constant is defined without the theory that commits to defining its meaning. The Newtonian inertial electron mass is a problematic element of Newtonian theory. Avoiding its problems makes a theory not needing Planck's constant.

We base our approach on quantum mechanics defined without a scale for amplitudes. Observables are given by $<\hat A> = <  \psi | \hat A|\psi>/<  \psi | \psi>$ in which the scale of the wave function cancels out\footnote{The expression using a density matrix $\rho$ is $<\hat A> =tr(\rho \hat A)/tr(\rho)$.}. It is conventional to normalize the wave function, and reduce the ambiguity to an unobservable overall phase, but mathematicians recognize the wave function as a projective or ray representation: producing a symmetry we call ``quantum homogeneity,'' which is admittedly not that new. Given a formula for the action $S$, the principle is \ba \delta S =0, \label{action} \ea whereby $S$ is dimensionless. The number $0$ on the right hand side of Eq. \ref{action} is special. The attempt to give $S$ and $0$ dimensions, a scaling $0 \ra \lambda 0$, is unobservable. If $S$ could be given a meaningful scale, it would violate quantum homogeneity. Eq. \ref{symmy} in Section \ref{sec:pathint} expresses the repercussions of this symmetry for $\hbar$ and $m_{e}$.

At this point the question of ``what is meant by quantum mechanics'' enters, because some will insist on giving action dimensions in their theory. Almost a century after discovery, there remain too many possibilities to resolve what quantum mechanics might be. In order not to offend anyone's definition, we set up our own system. It is based largely on dispensing with pre-history dominated by Newtonian work, energy, and the $MKS$ unit system that so sorely needed Planck's constant. 

Due to that history there are different default meanings of ``action.'' In Eq. \ref{action} the symbol $S$ {\it does not} refer to a Newtonian expression. It refers to the Action of the quantum mechanical wave function, and quantum fields, as explained below. For emphasis, we completely abandon the idea that the {\it classical action} $S_{cl}$ {\it in classical Newtonian units} has any predictive power for 21st century formulation of fundamental theory. (Once again this should not be controversial, but to avoid backlash we will call it ``our theory.'') In Eq. \ref{action} the symbol $S$ refers to the complete action of the Universe, which we believe has no absolute scale.

When Eq. \ref{action} is taken seriously the Hamiltonian $H = -\p S/\p t$ has dimensions of $time^{-1}$. It happens that measurements of time and frequency are by far the most accurate in current experimental physics. We believe this is not a technological accident. It seems to be something fundamental, and inherent in quantum theory; it certainly emerged when technology became mature and self-consistent with quantum theory. In recognition of dimensionless action, our primary dynamical equation computes the time evolution of the wave function without introducing extraneous units: \ba \i \dot \psi = \hat \Omega  \psi 
\label{freqq} \ea Here $\hat \Omega$ is the {\it frequency operator}. The equation computes frequencies in just the same units where frequencies are measured. 

For example, the excitations of the Hydrogen atom were first observed in frequency (wave number) units. There is no need to convert the observed frequency to Newtonian ``energy'' and back to predict frequencies. The intermediate
step known as ``Planck's constant'' is avoided. If not obvious at this stage, any physics which can avoid a fundamental constant is new and worth examining. Dropping redundant constants at the first stage cleans up an unlimited number of relations downstream. For example: The Planck spectrum of a black body is naturally a function of frequency with a single frequency parameter. The frequency-temperature does not need to be converted to temperature in degrees Kelvin, nor do degrees Kelvin need conversion to Newtonian ``energy'' and back to completely describe the spectrum. Yet multiple conversions of units are a fact of the standard Planck spectrum, conventionally expressed with a term going like $( exp(\hbar \omega/k_{B}T)-1)^{-1}$. 

 
Quantum theory came out of thermodynamics and its definition of energy. When the general proportionality of energy and frequency was discovered, it was associated with whole-numbers of ``quanta'' also found to exist, leading to the term ``quantum mechanics.'' Yet now ``quantum mechanics'' is a misnomer. Whole number units of physical quantities are not matters of principle. They are dynamical facts predicted by dynamical equations, when and if they occur. That was supposed to replace and eliminate the ``unit of action'' postulated to mock up quanta using pre-quantum Newtonian physics. Yet the cult of the quantum of action remains strong. It seems to be an inherent feature of what is commonly called quantum theory. In comparison, we find faults in the Newtonian precepts that led to a quantum of action in the first place. Newtonian {\it inertial mass}, set up by convention in certain units, is the culprit and something we can do without in a fundamental theory.  

\subsubsection{Avoidable Constants} 

Early physics was full of arbitrary units for seemingly unrelated quantities that were later united by a theory. Energy, temperature, and time evolution at first were unrelated. The number known as Boltzmann constant $k_{B}$ had a different status in 1900 than today.  Planck's writings show he believed it was fundamental. He treated both $h$ and $k_{B}$ as {\it universal constants} with great symmetry, including both in setting up his system of``fundamental'' units. At the time ``temperature'' as the average energy of atoms was a {\it hypothesis}, and energy itself was not absolutely established as more fundamental. The interpretation of Boltzmann's constant has evolved. It is uncontroversial to call Boltzmann's constant an {\it avoidable} constant of unit conversion.

Avoidable constants represent human convention. Avoidable constants can be measured: one can measure the number of centimeters in an inch by comparing two rulers. However avoidable constants cannot be measured until a convention is imposed to define them: they are ``absolutely unobservable,'' but conditionally observable by introducing outside conventions for definition. There is nothing incorrect in setting up an arbitrary scale for action based on a reproducible subsystem, so long as it is recognized as arbitrary. The arbitrariness of units of energy and time are recognized by everyone, while action got a special dispensation. It appears this comes from circularly defining canonical transformations to exclude a phase-space scale transformation. In much the same way that general relativity allows scale transformations on space-time, because they can't be prohibited, our approach to quantum mechanics is unchanged when the scale of action is globally revised. It is a minor change to permit a global transformation, but there are consequences. 

One consequence of avoiding redundant physical constants is that other constants can be determined better. For example: the fine structure constant is conventionally defined by $\a =e^2/hc$, where (in history) the electric charge $e$, the speed of light $c$ and Planck's constant $h$ had previously been identified. When $\a$ is given that definition, the experimental errors in its determination are inherited from $e$, $h$ and $c$. Are any of these constants avoidable? That depends on the theory.

In the theory where $c$ is an absolute constant of free space we agree that $c$ is observable. Yet the number for $c$ depends on the units of length and time. By sensible agreement decades ago, the distance scale of the $meter$ was eliminated in terms of a reference value of $c$ and a frequency standard. It is important that theory played a role. If the theoretical symmetries of special relativity had not been accepted, the errors of the Internatlonal Prototype Meter (a platinum bar) would now dominate physics.

With $c$ eliminated the errors in $h$ currently dominate errors in $\a$. In our approach $\a$ comes directly from data, without intermediate definitions of $h$ or $e$. Then from data we can determine $\a$ more precisely. It also turns out that the value of $e$ is an avoidable fiction based on concepts of point charges, Coulomb's Law, and Newtonian physics that offer nothing fundamental. We avoid $e$ entirely, since it does not really occur anywhere: when used in calculations, it is shorthand for $\sqrt{\a}$. 

When the numerical value such as $c$ is reduced to a definition, something experimentally testable is given up. Such decisions, if well-motivated by theory, stand to be reviewed from time to time, and the topic of Lorentz symmetry violation happens to be very active. Whether or not one is interested in Lorentz violation, it is still the right decision to eliminate $c$. That is because fundamental constants must be used self-consistently. This is not always appreciated. The electron mass we mentioned might mean a Newtonian inertial parameter to one physicist, a number in Schroedinger's or Dirac's theory to others, and a feature of the Higgs coupling to someone else. None of these definitions are the same. Relating one usage to another needs a theory, which will involve the error bars and systematics of the theory.

I will show how to significantly improve the electron mass defined as a parameter in the fundamental Lagrangian. Our definition should not be controversial. Those familiar with perturbative renormalization know this is surprisingly intricate, and ultimately dependent on conventions. More surprisingly, the best current determinations of the ``electron mass'' continue to refer to the inertial mass of a {\it Newtonian electron} ! Planck's constant is again responsible for the fault. If one insists on measuring Newtonian electron masses in kilograms, the fundamental arbitrariness and inaccuracy of the kilogram enters both in the measurement and in Planck's constant. 


Giving up Planck's constant is a not mathematically challenging, but just as replacing $c$ by a constant represents a theory, it may seem momentous. We find it sharpens a vision of quantum mechanics itself, where many historical holdovers stand out more clearly and can be eliminated. Describing a theory without Planck's constant is quite a bit simpler than explaining every apparent paradox of the theory built to perpetuate Planck's constant. Because the historical elements dominate current thinking, much of the paper is concerned with reconsidering elementary issues from a new point of view. The paper is written so readers not concerned with the puzzles of conventional theory can turn straight to Section \ref{sec:precision}.

\subsubsection{Organization of the Paper}

Section \ref{sec:dynamics} presents a ``derivation'' of Schroedinger's dynamics along the lines of effective field theory. It is elementary, yet novel, avoiding the illusory ``golden road'' of predicting quantum mechanics from Newtonian physics. The new approach shows that $\hbar$ does not come from comparing quantum theory to data. Section \ref{sec:Energy} reviews the concept of mass and energy. Snippets of history known as ``modern physics'' are interesting to consider from the new viewpoint. How $\hbar$ came to have its special numerical value is explained in Section \ref{sec:enterK}. Measurement,``spin'', the uncertainty principle and several other issues are considered in Section \ref{sec:measure}. There is no perfect ordering of many possible topics, and a short discussion was felt preferable to an encyclopedic compendium answering every possible question. Section \ref{sec:gravity} considers whether the Planck scale of gravity involves Planck's constant. The fact that eliminating Planck's constant improves the determination of other constants is discussed in Section \ref{sec:precision}. 


\section{Dynamics}  
\label{sec:dynamics}  

The non-relativistic Schroedinger equation is often ``derived from'' Newtonian physics using substitutions rules involving Planck's constant. With respect for the founders, it might be considered absurd {\it a priori} in the 21st century to ``derive'' quantum physics, which is more fundamental, from classical physics, which is an approximation. A different derivation makes no reference to that starting point.  

\subsubsection{Derivation} 
Quantum systems have wave attributes. A generic linear wave equation for amplitude $\phi$ is \ba  
{\p^{2} \phi\over \p t^{2}} +c_{*}^{2} \NN^{2}\phi =W(\phi). \label{basic} \ea Symbol $c_{*}$ 
has dimensions of speed, and might depend on the system (label ``$*$''). The coupling $W(\phi)$ on the right hand side of Eq. \ref{basic} is often omitted in ``wave equations'', but 
something must couple waves to the world. The simplest possibility is a linear dependence: \ba W(\phi) = W(\vec x)\,  \phi (\vec x).\nn \ea 

The coupling $ W(\vec x)$ will be fit to experimental data, as we review momentarily. It is convenient to extract the constant part of $W$ from the spatially varying part $\tilde W(\vec x)$: \ba W(\vec x) = 
\omega_{*}^{2}+ \tilde W(\vec x). \label{ww} \ea Substitution gives  
\ba {\p^{2} \phi\over \p t^{2}} +c_{*}^{2} \NN^{2}\phi =\omega_{*}^{2}\phi+ \tilde W(\vec x)\phi. \nn
\ea A simple transformation removes the $
\omega_{*}^{2}$ term. Define \ba \phi = e^{-i \omega_{*}t} \psi, \label{subs} \ea where $
\psi$ is a symbol for $e^{i \omega_{*}t} \phi$. If $\phi$ was real-valued then $\psi$ becomes complex by an act of notation. In discussing spin and charge we will assume $\phi$ is complex from the beginning. Putting Eq. \ref{subs} into the wave Eq. \ref{basic} gives \ba { \p\phi
\over \p t} &=& e^{-i \omega_{*} t}( - i \omega_{*} \psi + {\p \psi \over \p t}); \nn \\
 { \p^{2 }\phi\over \p t^{2}} &=& e^{-i \omega_{*} t}( -\omega_{*}^{2}\psi- 2i   
\omega_{*} {\p \psi \over \p t} +{\p^{2} \psi \over \p t^{2} }). \label{det}  \ea  
The $\omega_{*}^{2}$ term cancels. We drop $\p^{2} \psi /\p t^{2} << 
\omega_{*} {\p \psi / \p t} $ for the low-frequency limit. The steps reduce the basic wave equation to \ba i  
{\p \psi \over  \p t} =- {c_{*}^{2} \over 2 \omega_{*}}\NN^{2}\psi + {\tilde W(\vec x ) \over 2 \omega_{*} } \psi. \nn \ea Simplifying the symbols gives \ba i{\p \psi \over  \p t} = \hat \Omega \psi. \label{eqmo} \ea Here $\hat \Omega$ is the {\it frequency operator}, in these approximations given by \ba \hat \Omega = - {c_{*} \lambda_{*} \over 2 }\NN^{2} + U(\vec x) . \nn \ea where\footnote{To simplify notation the symbol $\lambda_{*}= {  c_{e} /\omega_{*}}$ does not introduce a conventional $1/2 \pi$. } \ba \lambda_{*}= {  c_{*} \over \omega_{*}}; \:\:\: \:\:\:  U(x) = {\tilde W(\vec x ) \over  2 \omega_{*}} \nn \ea  

Symbol $U(x)$ will be called the ``interaction function''. In the next Section we will find the interaction function for electrons from experiment. While much of that will be familiar, it is important to be conceptually complete and independent. Rather than repeat historical arguments in the historical order, we will keep track of each source of information and its consequences for the quantum system without imposing any Newtonian prejudices. 

\paragraph{Comment:}Some will recognize our approach as a low-frequency effective field theory. It is possible to object that the quantum mechanical wave functions are not classical fields. This is true, but irrelevant. No matter how $\psi$ is interpreted, $\hbar$ is not needed to set up the dynamics. There is a related issue from old probabilistic arguments introducing $\hbar$ early, and reasoning on its basis to supposedly predict elements of the dynamics. It is evidently not needed, because we've not used it.

Heisenberg's matrix mechanics predated Schroedinger's equation via a presentation that made Planck's constant appear indispensable. For our purposes it is only necessary to define the dynamics once. {\it Given} the frequency equation without reference to $\hbar$, the corresponding Heisenberg operator equations of motion follow without $\hbar$. More discussion appears in Section \ref{sec:gravity}.

\subsection{Solutions and Data Fitting} 

\begin{figure}[htbn]
\begin{center}
\includegraphics[width=4in]{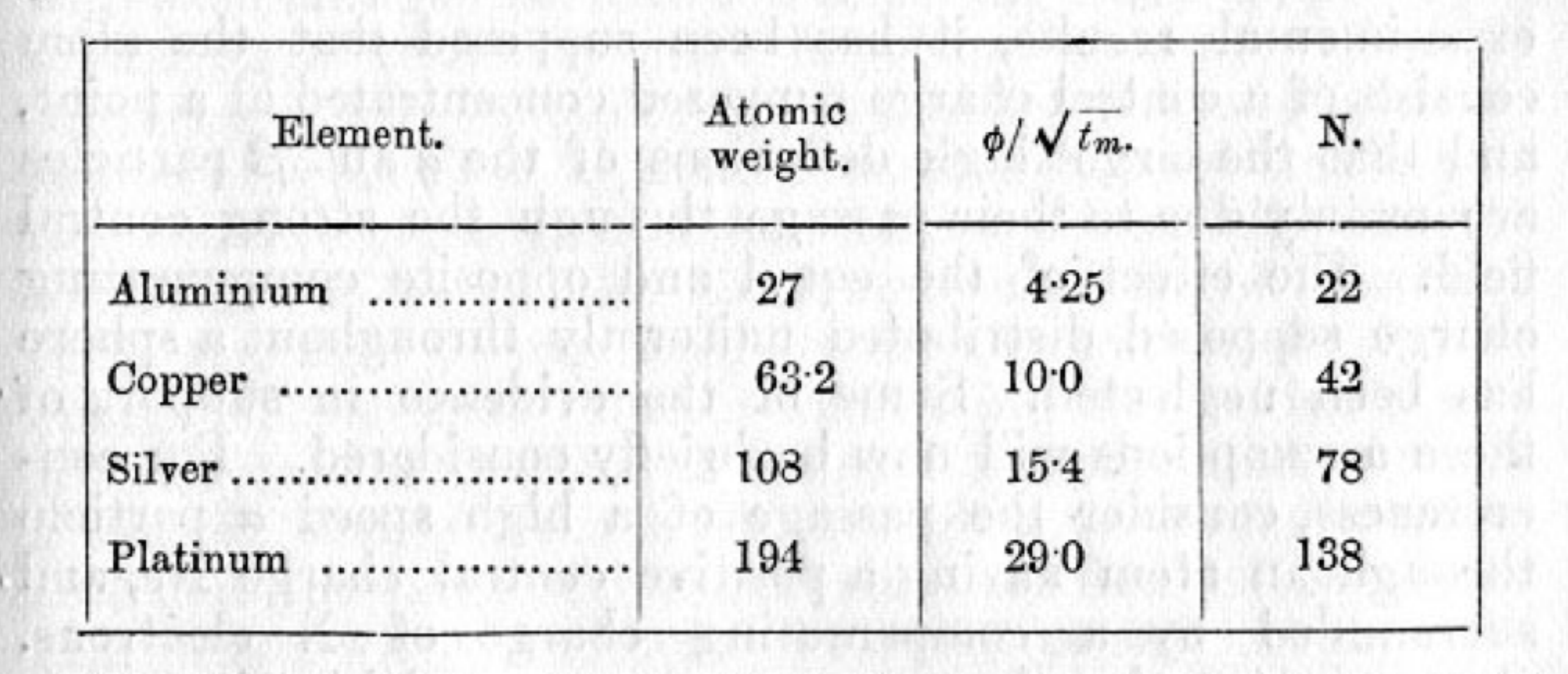}
\caption{J. A. Crowther's 1910 data on electron scattering from Rutherford's 1911 {\it Philosophical Magazine} article. The column $\phi/\sqrt{t}$ (in units of $radians \, cm^{-1/2}$) refers to the target thickness $t$ for which half the total flux was scattered to angle exceeding $\phi$. $N$ is Rutherford's computation of nuclear charge $Z$, whose relation to the atomic weight was not established at the time. The table yields a parameter $\hat \sigma \sim 1 \pm 0.3 \times 10^{-26} \, cm^{2}$.}
\label{fig:Ruthy.pdf}
\end{center}
\end{figure}
\label{sec:fitting}

We briefly discuss fitting the constants of the frequency equation to experimental data. We follow the unusual but self-consistent path of not introducing Newtonian assumptions. All the data necessary was available before 1925, but we also consider the 1927 work of Davison and Germer. \bit \im Rutherford interpreted the data of Geiger and Marsden's data on alpha-particles, and Crowther for beta particles, electrons. Rutherford's main contribution was a classical scattering model we see no reason to repeat. Rutherford fit an electron-atom cross section given by \ba {d \sigma \over d \Omega} =({ c \over v_{0}})^{4}Z^{2}{ \hat \sigma \over sin^{4}( \theta/2)}. \label{ruthmodel} \ea From basic {\it wave mechanics} of our scattering theory in Born approximation\footnote{The scattering equation can actually be solved exactly for the same interaction function, eliminating the Born approximation.} the $sin^{4}( \theta/2)$ angular distribution {\it predicts the interaction function} going like $1/r$: \ba U(\vec x) & = &   {\kappa Z  c_{e}\over r},  \label{Ufound} \ea where the constant $\kappa$ is dimensionless. There is tremendous information in the shape of the interaction function. Using the modern values of $Z$, we fit the 1910 data (Fig. \ref{fig:Ruthy.pdf}) to find $\hat \sigma =(1.03  \pm 0.28) \times 10^{-26} \, cm^{2}.$ The error is statistical; Crowther's later data indicates this has systematic errors of relative order 50-100\%. The number implies \ba \kappa & \sim  & ({ 2 \times 10^{-13} cm \over   \lambda_{e}} )\, (  {c\over c_{e}})^{2}   . \label{kappafound} \ea One combination of constants has been determined. 

\im Bound states and characteristic frequencies (normal modes) are found by solving \ba \hat \Omega \psi_{n} = \omega_{n} \psi_{n} . \nn \ea Wavelengths and frequencies are the observables of spectroscopic experiments. The wave equation predicts zero current and no radiation from frequency eigenstates, resolving a pre-quantum puzzle. When there is a current, it predicts radiation frequencies that are differences of electron eigenfrequencies $\Delta \omega_{n_{1}n_{2}} =\omega_{n_{1}}-\omega_{n_{2}} $. 

\begin{figure}[htbn]
\begin{center}
\includegraphics[width=3in]{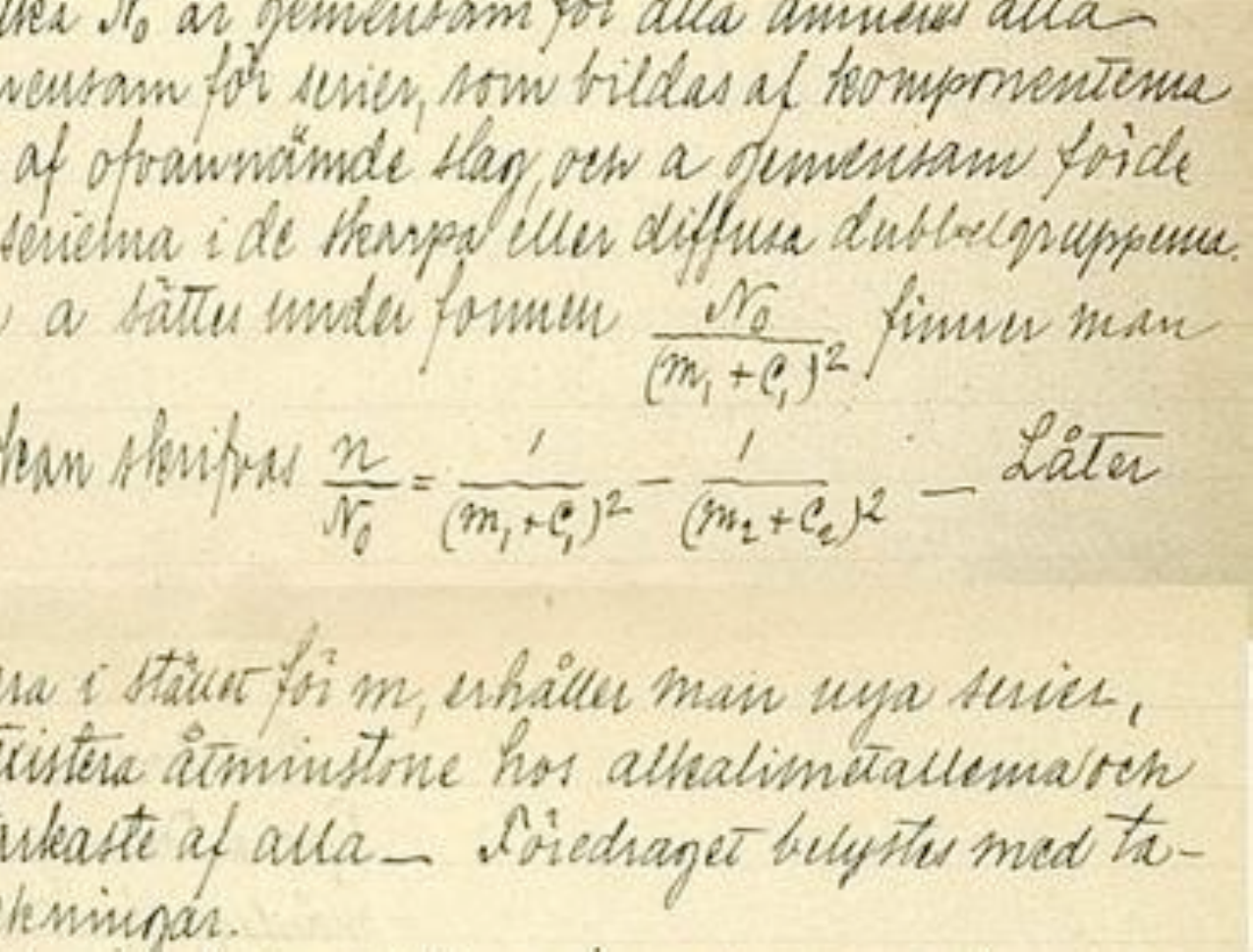}
\caption{ \small Excerpt from the minutes of Society for Physics and Mathematics (Lund, Sweden) of November 5th 1888, showing Rydberg's formula. From Lund University Physics\cite{rydberg}.}
\label{fig:Rydberg}
\end{center}
\end{figure}

Rydberg observed for the Hydrogen atom \ba \Delta \omega_{n_{1}n_{2}} = 2.07 \times 10^{16} ({1 \over n_{1}^{2}}- {1\over 
n_{2}^{2}}) \,s^{-1}.  \label{balmer} \ea Here $n_{1} \geq 1$, $n_{2} \geq 1$ are integers\footnote{ Rydberg's formula included a ``quantum defect'' $n_{i} \ra n_{i}+ c_{i}$, improving the precision, which Bohr omitted. Strictly speaking, Rydberg's fit reported\cite{rydberg} a {\it wave number} $\omega_{H}/2 \pi c= 10, 972, 160 \, m^{-1}$, equivalent to the {\it frequency scale} $\omega_{H}=2.07 \times 10^{16} \,s^{-1}$ cited.}. It is very significant that Rydberg's frequency spectrum is consistent with the prediction from the scattering interaction, so that the interaction function has been determined twice. The data for $\omega_{H}$ fixes one new combination of parameters: \ba  {\kappa^{2}  c_{e}  \over 2 \lambda_{e} } = 2.07 \times 10^{16} \,s^{-1}.\label{valone} \ea 

\begin{figure}[htbn]
\begin{center}
\includegraphics[width=3in]{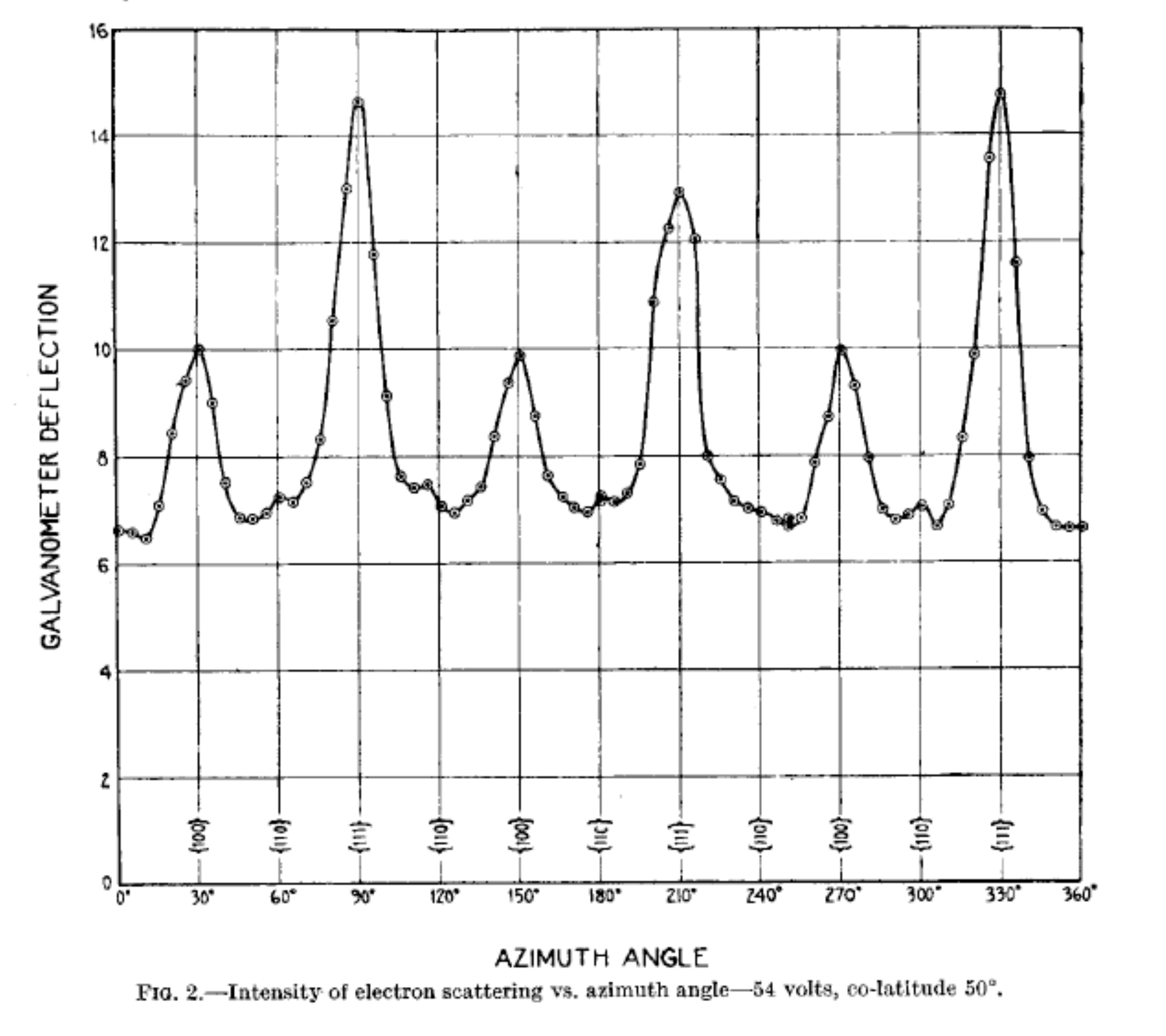}
\caption{ \small Davisson and Germer's data for electron intensities scattered from Nickel 
as a function of azimuthal angle relative to the beam at fixed polar angle $\theta=50^{o}$.  }
\label{fig:DavissonGermerFig2.pdf}
\end{center}
\end{figure}

\im In 1927 Davisson and Germer\cite{davisson} observed electrons scattering from Nickel crystals. They used the crystals as diffraction gratings and measured the scattered angular distribution of electrons. Our fit to Fig. \ref{fig:DavissonGermerFig2.pdf} gives and observed wavelength $\lambda =1.65 \times 10^{-8}$ cm. For this measurement the $DG$ beam had a frequency\footnote{Chemical batteries were long the standard for electrical measurement. About 9 $AA$ batteries in series will ionize Hydrogen. $DG$ used the equivalent of 54. } $\omega = 3.97 \omega_{H}$. The experiment fixes one parameter in the free-space dispersion relation, \ba  c_{e} \lambda_{e}  = 1.14 cm^{2}/s .\nn \ea \im The wave model predicts a characteristic length scale $a_{0} =\lambda_{e}/\kappa $ for the ground state of Hydrogen. By 1899 Dewar had already produced solid Hydrogen. Its density of 0.07 $gm/cm^{3}$ at the boiling point is reported in 1904, ``Physical Constants at Low Temperatures. (1)--The Densities of Solid Oxygen, Nitrogen, Hydrogen, etc''\cite{Dewar1904}. Estimating $r_{H}\sin 3 a_{0}$ and atomic volume $4 \pi r_{H}^{3}/3$ gives $r_{H} \sim 2.88 \times 10^{-8} \, cm$. Then \ba { \lambda_{e} \over \kappa } \sim { 2.88 \times 10^{-8} \over 3} \sim 10^{-8} \, cm. \label{size} \ea This is a relatively crude estimate, but no harm is done by including it.

\begin{figure}[htbn]
\begin{center}
\includegraphics[width=2.8in]{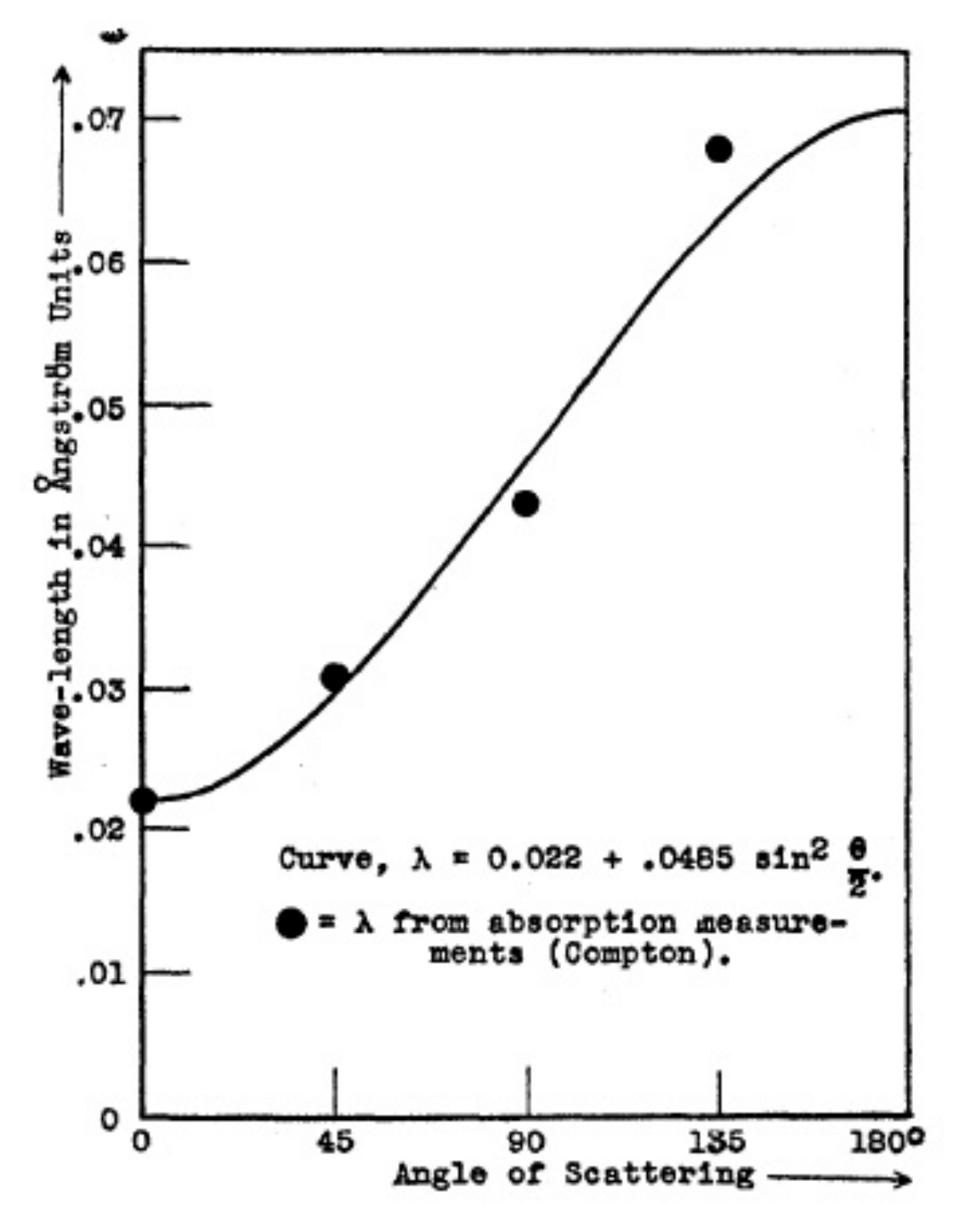}
\caption{ \small Compton's data from the 1923 paper. The legend cites the observable scale $0.0485 \, A^{0} =4 \pi \lambda_{e}$. }
\label{fig:ComptonData}
\end{center}
\end{figure}
 
\im Compton's 1923 paper\cite{compton} "A Quantum Theory of the Scattering of X-Rays by Light Elements" treated electrons and photons as relativistic particles conserving energy and momentum. Compton's early interpretation continues to be cited as a proof of both a particle interpretation and the necessity of Planck's constant. Our approach uses neither because they add no information. Take a moment to review the calculation. 

From Lorentz covariant perturbation theory, the differential cross section $d \sigma$ is given in standard form by \ba d \sigma ={d^{3}k_{1}' \over (2 \pi)^{3}2 k_{1}^{'0}}...{d^{3}k_{n}' \over (2 \pi)^{3}2 k_{n}^{'0}}(2 \pi)^{4}\delta^{4}(k_{1}+k_{2}-\sum_{f} k_{f}') |M|^{2}, \nn \ea where $M$ is a Lorentz-invariant scattering amplitude. Symbols $k^{\mu}=(\omega/c, \, \vec k)$ are the frequencies and wave numbers of participating waves. The delta function comes from translational invariance, and gives \ba k_{e}+k_{\gamma}  &=&  k_{e}'+k_{\gamma}' ; \nn \\ k_{\gamma}\cdot k_{e} &=& k_{\gamma}'\cdot k_{e}'= k_{\gamma}'\cdot (k_{e}+k_{\gamma});  \nn \\ (k_{\gamma} - k_{\gamma}') \cdot k_{e}  &=& k_{\gamma}'\cdot k_{\gamma}. \label{kine} \ea The electron wave dispersion relation is \ba ({\omega_{e} \over c})^{2}-\vec k_{e}^{2}-1/\lambda_{e}^{2}=0. \nn \ea  In the rest frame the electron wave vibrates with $\vec k=0$ and frequency $c/\lambda_{e}$. Evaluating Eq. \ref{kine} gives \ba ( {\omega_{\gamma} \over c} - {\omega_{\gamma}' \over c}  ){c \over \lambda_{e} } = {\omega_{\gamma} \omega_{\gamma}' \over c^{2}}(1-cos\theta). \nn \ea Dividing both sides by $ \omega_{\gamma} \omega_{\gamma}' $ gives the change in the photon wavelength $\Delta \lambda_{\gamma}$: \ba \Delta \lambda_{\gamma}= 4 \pi \lambda_{e} \, sin^{2}(\theta/2). \nn \ea While Compton need extraneous constants to make the calculation, none appear in this approach.

Fig. \ref{fig:ComptonData} shows Compton's data from the original paper, and the fit: \ba \Delta \lambda_{\gamma} = 0.485 A^{0} sin^{2}(\theta/2). \nn \ea Compton's data gives \ba  4 \pi \lambda_{e} =4.85 \times 10^{-11}\, cm; \nn \\ \lambda_{e} = 3.86 \times 10^{-11} \, cm. \label{lambdae}\ea  \eit

\subsection{Summary of Preliminary Fits} 

\begin{table}[htbn] 
  \centering 
\resizebox {.9\textwidth }{!}{

  \begin{tabular}{cccc}
\hline
 $Source$ &  $Data$  &  Relation   \\
   \hline \\ 
 $ Crowther/Rutherford $  &  Electron-Atom Scattering   & $({c_{e}/c})^{2} \kappa \lambda_{e} \sim  2 \times 10^{-11} \,cm$     \\
 $ Rydberg $ & Hydrogen Frequency Spectrum & $ \kappa^{2} c_{2} \lambda_{e} \sim  4.14 \times 10^{16}s^{-1} $ \\ 
  $Davisson \, and \,Germer $ & Electron-Nickel  \,Scattering  & $  c_{e} \lambda_{e} \sim 1.14 cm^{2}/s$  \\ 
 $Dewar $ & Solid Hydrogen Density & $\lambda_{e}/\kappa \sim 10^{-8} \, cm$  \\ 
  $Compton $ &  Electron-Light Scattering & $\lambda_{e}  \sim  3.86 \times 10^{-11} \, cm$. \\
 
\end{tabular} }

  \caption{ \small Summary of experiments and parameter relations in a preliminary fit to the quantum theory, as discussed in Section \ref{sec:fitting}. }
  \label{tab:fits}
\end{table}

\begin{figure}[htbn]
\begin{center}
\includegraphics[width=6in]{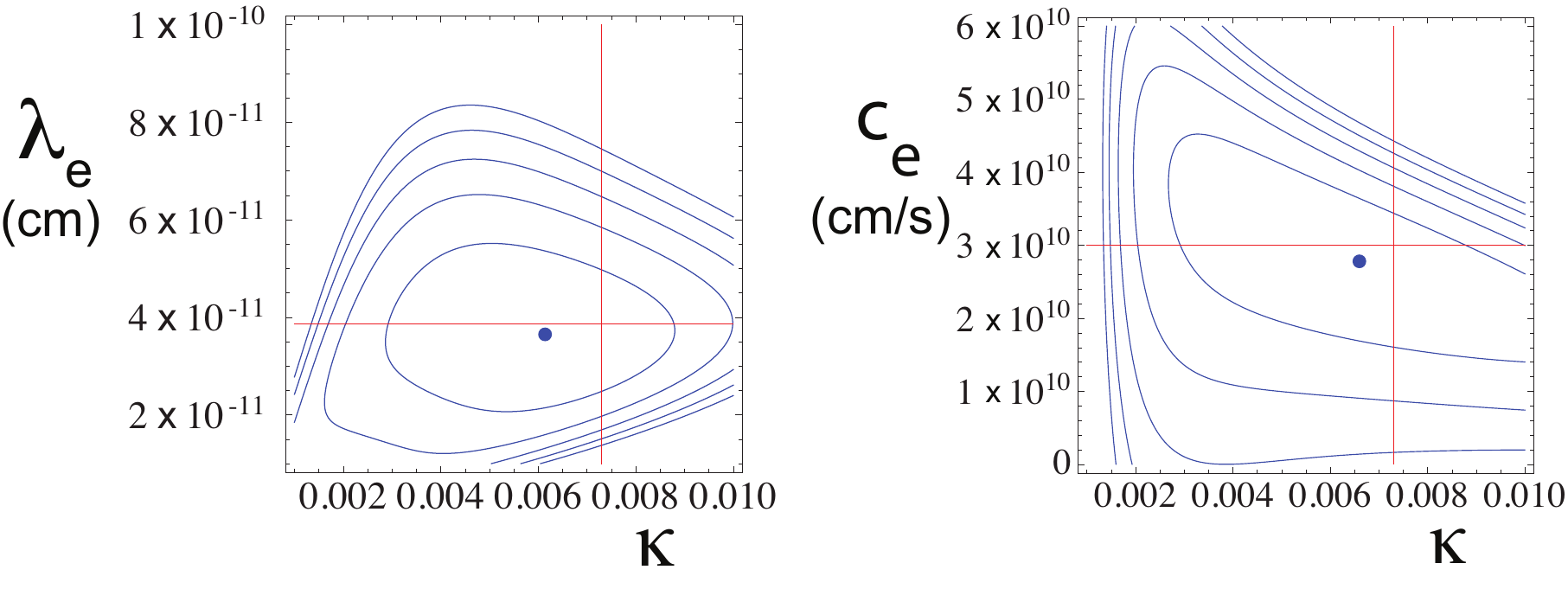}
\caption{\small Contours of $\chi^{2}$, the summed-squared differences of data versus fit obtained from the independent relations given in Table \ref{tab:fits}. {\it Left panel}: As a function of parameters $\kappa$ and 
$\lambda_{e} $ with $c_{e} \ra c$. {\it Right panel}: As a function of parameters $\kappa$ and $c_{e}$ with 
$\lambda_{e}$ given by Compton's value. Dots shows the points of minimum $\chi^{2} \sim 0.24$ in both cases. Contours are $\chi^{2}=1$, 2, 3... Lines show modern values $c=3\times 10^{10}cm/s, \, \lambda_{e}=3.87 \times 10^{-11}, \, \kappa =1/137$ all lie inside the range of $\chi^{2} \lesssim 1$. }
\label{fig:AlphaLamCFit.eps} 
\end{center}
\end{figure}

Table \ref{tab:fits} summarizes the experiments and parameter relations each implies. Since early data seldom included error bars, no great effort has been made to assess errors in this preliminary exercise. Each of the independent relations given in Table \ref{tab:fits} predicts a relation of parameters $f_{j}( \kappa , \, c_{e})=constant_{j} $ We define $\chi^{2} =\sum_{j} \, (f_{j}( \kappa , \, c_{e})/constant_{j}-1 )^{2}$, in order to scale out the absolute size of numbers and units. While $\chi^{2}$ is not weighted by errors, which are unavailable, it can be rescaled by any uniform error estimate $\sigma^{2}$. Figure \ref{fig:AlphaLamCFit.eps} shows two contour plots of $\chi^{2}( \kappa, \, \lambda_{e} )$ with $c_{e}\ra c$ (left panel) and $\chi^{2}( \kappa, \, c_{e} )$ with $\lambda_{e} \ra Compton's \, value$ (right panel). 

The best fit value $\chi^{2} \sim 0.24$ comes in both cases, with preliminary values of $\kappa $ and $c_{e}$: \ba 
  \kappa &=&  0.0061= {1 \over 164} ;\:\:\:\:\:  \lambda_{e}=3.65 \times \,10^{-11} \, cm \:\:\:\:\: (c_{e}\ra 3\times 10^{10} \, cm/s)  \nn \\
 \kappa &=& 0.0066 = {1 \over 151}; \:\:\:\:\:  c_{e} = 2.78 \times 10^{10} \, cm/s :\:\:\:\: (\lambda_{e}\ra 3.86 \times 10^{-11}\, cm) . \nn \ea 
These results are very acceptable compared to modern values. Disregarding all the data in favor of Compton's value of $\lambda_{e}$ and Rydberg's determination of $\Omega_{H}$, which are probably more reliable yields a one-parameter relation \ba \kappa^{2}c_{e} =1.59 \times 10^{6} \, cm/s \nn \ea Assuming $c_{e} \ra c$ as indicated by the rest of the data gives \ba \kappa =\sqrt{{ 1.59 \times 10^{6} \, cm/s \over 3 \times 10^{10} cm/s   }} = 0.0073 ={ 1\over 137.0}. \nn \ea 

Our interaction constant $\kappa$ came to be called the {\it fine structure constant,} symbol $\a$, which we adopt: \ba \a \equiv \kappa \sim {1 \over 137}. \nn \ea It is important that $\a$ comes directly from the quantum data, without using extraneous definitions in terms of the classically-motivated concepts of $e$ and $h$.

\subsubsection{Where is Planck's Constant?} 

The development shows quantum mechanics can be developed without Planck's constant. It is absent in a theory developed directly from quantum mechanical data. 
 
Each quantum system, such as the electron, muon, proton... is characterized by a frequency scale $\omega_{*}$ or equivalent length scale $\lambda_{*}$. A few of these fundamental constants are given in Table \ref{tab:1}. They can be found from data by ratios. For instance, a simple comparison of muonium to positronium level frequencies will give $\omega_{\mu}/\omega_{e} \sim 207$.

\begin{table}[!]
  \centering 
\begin{tabular}{ccc}
\hline \\

$System $  &  $\omega_{*}$ ($s^{-1}$)  & $\lambda_{*} $ ($m$) \\ \\
\hline \\
 $e$  &  7.79 $\times  10^{20} $ &  $ 3.862 \times 10^{-13}$   \\
 $\mu$   &  1.604 $\times  10^{23} $  &  $ 1.869  \times 10^{-15}$  \\
  $\tau$   & 2.698 $\times  10^{24} $ & $1.111 \times 10^{-16}$    \\
    $p$   & 1.43 $\times  10^{24} $ &  $2.104 \times 10^{-16}$    \\
      $n$   & 1.427 $\times  10^{24} $   &  $2.101 \times 10^{-16}$    \\
 $\gamma$  &  0 &  $\infty $       \\
$W^{\pm}$  &  1.221 $\times  10^{26} $ & $ 2.455 \times 10^{-18}$ \\
$Z^{0}$  &  1.385 $\times  10^{26} $ & $ 2.165 \times 10^{-18}  $  \\ \\ \hline \end{tabular}

  \caption{ \small Characteristic frequency ($\omega_{*}$) and length ($\lambda_{*} $ ) scales of selected physical systems. }
  
  \label{tab:1}
\end{table}

The electromagnetic interaction is characterized by a pure number $\a \sim 1/137$. The strong and weak interactions have similar dimensionless parameters. It is impossible to compute Planck's constant from these numbers. 

\subsubsection{Non-Relativistic Action} 

As conventional, our quantum dynamics (Eq. \ref{eqmo}) can be expressed using an action principle. It is the outcome of varying the action $S_{QM}$: \ba S_{QM}=\int dt \, L_{QM} =\int dt  \int d^{3}x \, i \psi^{*} \dot \psi - \psi^{*} \hat \Omega \psi. \label{act} \ea Variation by $\psi^{*}$ gives \ba \i \dot \psi - \hat \Omega  \psi 
=0,  \label{eq2} \ea from which the frequency operator $ \hat \Omega $ is identified. After solving the equations the numerical value of the action can be computed: \ba S_{QM}= 0. \nn \ea 

Observe that the Lagrangian $L_{QM}$ has the form \ba L_{QM} =\sum_{x} \, p_{x}\dot q_{x}-H(q_{x}, \, p_{x}),\nn \ea where \ba q_{x} &=& \psi(x); \nn \\  p_{x}  &=& {\delta  L_{QM} \over \delta  \dot q_{x}} = i  \psi(x)^{*}. \label{momenta} \ea Here $\sum_{x}$ is discrete notation for the volume integral. The fact that $i \,\psi^{*}$ are the canonical momenta of the field provides a physical interpretation of observables $<\psi|\hat A|\psi> =\sum_{x} \, p_{x}\hat A_{xx'} q_{x'}$. This is a projective map {\it from} many canonical variables {\it into} certain canonical quantities with definite transformation properties. Neither probability interpretation nor semi-classical arguments involving Planck's constant are needed to explore this. 

For example, a bulk translation of the coordinate system is defined by \ba \psi(x) \ra \psi_{\vec a}(\vec x) = \psi(\vec x-\vec a). \nn \ea  Noether's theorem using $L_{QM}$ (Eq. \ref{act}) then predicts the conjugate total momentum: \ba \vec P = \int d^{3}x \, \psi ^{*} (-i   \NN )\psi =<-i \NN> \label{momdef}.  \ea Since $<\psi|\psi>$ drops out of observables we normalized it to one. By using Noether's theorem Eq. \ref{momdef} reproduces a familiar step of traditional theory involving heuristic (and actually redundant) quantum postulates.

Since $\vec P$ are full-fledged canonical variables, there are conjugate variables $\vec Q$, which we represent with an projective map involving operator $\vec{\cal Q}$, and its transformation properties: \ba \vec Q = d^{3}x \, \psi ^{*}\vec{\cal Q} \psi ; \nn \\  \vec Q \ra \vec Q+\vec a. \nn \ea  The basic test of conjugacy lies in the Poisson bracket: \ba \{ Q_{i}, \, P_{j} \}_{PB} =-i\sum_{x} \, \left( {\delta Q_{i} \over \delta \psi_{x}}{\delta P_{j} \over \delta \psi_{x}^{*}}-   {\delta P_{j} \over \delta \psi_{x}}{\delta Q_{i} \over \delta \psi_{x}^{*}} \right). \nn \ea Computing the derivatives gives \ba \{ Q_{i}, \, P_{j} \}_{PB} =-i \sum_{x} \psi_{x}^{*} [ {\cal Q}_{i}, \,  {\cal P}_{j} ]\psi_{x} .\nn \ea With $ <\psi|\psi>=1$ we obtain the map between the operator algebra and Poisson bracket \ba \{ Q_{i}, \, P_{j} \}_{PB} =\delta_{ij} \ra  [ {\cal Q}_{i}, \,  {\cal P}_{j} ]=i \delta_{ij}   \label{commute} \ea 

That is, Noether's theorem and group representations predict the semi-classical operator substitution rules postulated early in quantum mechanics are kinematic consequences of the wave theory\cite{john}. The reason our approach does not need Planck's constant, while the historical one did, lies in accepting an infinite number of degrees of freedom for the ``electron'' in the first step. It is an experimental fact of quantum waves. In order to obtain a constant $\lambda$ with any desired dimensions on the right hand side of Eq. \ref{commute}, it is sufficient to multiply the action (Eq. \ref{act}) by the same constant, which has no consequences whether $\lambda=\hbar$ or any other value.


\subsubsection{Path Integrals and Special Relativity}
\label{sec:pathint}

Path integrals are often cited as a starting point for quantum field theory. One might ask whether quantum field theory, which is a highly comprehensive approach, might more fundamental than basic quantum mechanics, even though developed as a generalization. It is possible to question whether field theory or relativity somehow puts Planck's constant into the theory. 

Then consider the action {\it in engineering units} for relativistic quantum electrodynamics, which is\footnote{Units $c=1$ used here should not cause confusion.} \ba S_{QED} =\int d^{4}x \, \bar \psi( i \hbar \slasha {\partial }- e \slasha A -m_{e}) \psi - {1 \over 4}F_{\mu \nu}F^{\mu \nu}. \label{QEDact} \ea Here $\psi$ is a Dirac field, $m_{e}$ is a constant called the Lagrangian mass parameter, and $F_{\mu \nu}$ is the electromagnetic field strength tensor. Now to wrongly ``prove'' that Planck's constant is involved, each field configuration in the path integral is weighted by $exp(i S_{QED}/\hbar)$, in which we see $\hbar$ explicitly. 

The claim is false, because the action of Eq. \ref{QEDact} was multiplied by $\hbar$ when formulated by historical substitution rules. The constant naturally cancels out in expressions using $S/\hbar$. In more detail, the path integral is invariant under a change of measure. Let $A_{\mu}' =  eA_{\mu}/\hbar$. At each point in space-time, let \ba d [A ] d [\psi ] d [\bar \psi ] &=&  d[\Phi] ; \nn \\   d[\hbar A'/e ]d [\psi ] d [\bar \psi ] &=&   d[\Phi]'. \nn \ea Observable correlations are represented by \ba <O(A, \, \psi, \, \bar \psi) > &=& {\int d[\Phi] \,  e^{i S_{QED}/\hbar}O(A, \, \psi, \, \bar \psi )\over  \int d[\Phi] \, e^{i S_{QED}/\hbar   } }, \nn \\ &=&  \nn {\int d[\Phi]' \,e^{i S_{QED}/\hbar   } O(\hbar A'/e, \, \psi, \, \bar \psi )\over  \int d[\Phi]' \, d [\bar \psi ] e^{i S_{QED}/\hbar   } }\ea The action exponent, including division by $\hbar$, transforms to \ba S_{QED}/\hbar =\int d^{4}x \, \bar \psi( i \slasha \partial -   \slasha A' - \omega_{e}) \psi - {1 \over 4 \a}  F_{\mu \nu}'F^{\mu \nu '},\label{qed} \ea where \ba \omega_{e} ={m_{e} \over \hbar} ;\:\:\:\:  \a = {e^{2} \over \hbar  } .\label{egone} \ea  Notice the electric charge $e$ itself does not appear in Eq. \ref{qed}. The two parameters $\omega_{e}$ (or $\lambda_{e}$) and $\a$ determine the path integral just as in our approach to quantum mechanics. 

Since $\hbar$ cancelled out, it cannot be measured in experiments using Eq. \ref{qed}. However nothing stops one from keeping it in the formalism while respecting a parameter symmetry. When one uses three parameters to represent two constants, there is one symmetry: \ba \hbar \ra \xi \hbar; \:\:\:\:\:  m_{e} & \ra & \xi m_{e}; \nn \\ {\hbar \over m_{e}} & \ra & {\hbar \over m_{e}} ={1\over \omega_{e}}\:\:\: invariant. \label{symmy} \ea

\section{What Do We Mean by Energy and Mass?} 
\label{sec:Energy}

\subsubsection{Newtonian Inertial Mass}

The ``mass'' of Newton's era was a ``measure of the quantity of matter.'' It is strongly tied up with intuitive, post-medieval expressions of ``force'' as a primary concept. The tradition constantly cites simplistic ideas of {\it additivity} of mass and force, and probably cannot proceed without them. We claim those concept of are no longer a starting point for fundamental physics. They are irrelevant in general, but relevant here to find out how $\hbar$ crept into physics early.

The Hamiltonian describing Newton's world is \ba H_{N} =\sum_{i} \,{\vec p_{Ni}^{2} \over 2 m_{Ni}}+ \sum_{i j} \, V_{ij}(|\vec q_{i}-\vec q_{i}|)\nn \ea This reproduces Newton's three ``Laws'' of motion in Newton's coordinates. Since it is not fundamental, we do not find a high obligation to reproduce the model from fundamental physics. But already by making a Hamiltonian the starting point, the theory has been revised, and the interpretation of the constants of the theory revised.

Given the Newtonian mass as a parameter in the action, its meaning and units are derived from the action. On the naive basis that $\delta S=0$, let us explore classical physics of dimensionless action. By inspection of this {\it macroscopic} action we have units \ba  q_{i} & \ra &  meters; \nn \\ p_{Ni} & \ra &  meters^{-1}; \nn \\ H_{N}   & \ra &  seconds^{-1}; \nn \\ mass_{N} & \ra &  seconds/meter^{2}. \nn \ea  

This defines the $M \slasha K S$ system where the kilogram ($K$) units of mass is never introduced by external standards. Can one do classical physics this way? Force is the gradient of the Hamiltonian: \ba \vec F_{N} = {d \vec p_{N} \over dt} =- \NN H_{N}. \nn \ea Force has units of $(meter \cdot seconds)^{-1}$. Consider a force of (say) 3 inverse $meter-seconds$ in the $x$ direction, and apply it to a mass of (say) 5 {\it seconds per square-meter}. Compute the acceleration \ba a_{x} ={F_{Nx} \over m_{N}} =3 (meter \cdot seconds)^{-1} \, {1\over 5} \, { meter^{2} \over  seconds} = {3 \over 5}{ meters \over seconds ^{2}}. \nn \ea 

Glue together two masses of $m_{1}=5 \, seconds/meter^{2}$ and $m_{2}~=10 \, seconds/meter^{2}$. What makes us think their inertial masses add? Go to center of mass ($cm$) and relative coordinates. Neglecting deformation of the glue, the relative coordinate drops out. Hamilton's equations tell us the $cm$ coordinate has effective mass $m_{1}+m_{2}=15 \, seconds/meter^{2}$. Applying the same force as before (using ``spring balances'', etc) it will accelerate at 3/15 $meters/seconds^{2}$. Proceeding this way we can re-build classical physics, including the theory of work, thermodynamics, steam-engines, entropy and the Planck distribution, without ever introducing the kilogram. 


There is a different theory where the mass parameter $m_{R}$ naturally has units of frequency. The ``free particle'' Hamiltonian in this theory is \ba  H_{R}=\sqrt{\vec p_{R}^{2}c^{2} + m_{R}^{2}}.\label{relham} \ea Since $m_{R}$ in Eq. \ref{relham} is a constant, it does not depend on the initial conditions of the theory. That produces one fact relating particles with energy and momentum $E_{R},  \, p_{R}$ to those with a different energy and momentum $E_{R}',  \, p_{R}',$ which is the Lorentz transformation. 

%
%
%

On dimensional grounds $m_{R}$  is simply a frequency scale for the energy, which has units of frequency. An explanation is needed why $m_{N}$ ever became popular. In the regime of small $p_{R}$, Eq. \ref{relham} becomes \ba H_{R} \sim {\vec p_{R}^{2}c^{2} \over 2m_{R}}. \nn \ea If the two theories describe the same thing, then
\ba m_{R} &=& m_{N} c^{2};  \nn  \\ seconds^{-1} &=&{seconds \over meters^{2}} \times {meters^{2} \over seconds^{2}}. \ea  

So far we've shown units are consistent, without going to the step of unit standardization. In both the Newtonian and relativistic $M\slasha K S$ systems it is sufficient to choose a certain lump of material (standard ``mass artifact''), and declare that it has one (1) unit of $seconds/meter^{2}$ of Newtonian mass. Such an object traveling at 1 $meter/second$ has an energy of $mv^{2}/2= (1/2) \, seconds^{-1}$, defining the energy unit. The same object has $m_{R}=9 \times 10^{16} \, seconds^{-1}$ of relativistic mass. Standardization is then tied to the mass artifact, which might have been a different lump of material, related by a simple scale factor. Different scale factors then translate to different units chosen for time.

Our quantum mechanical fits to electron waves have given $\lambda_{e}=3.65 \times \,10^{-11} \, cm$, or $\omega_{e} =c/\lambda_{e} =8.22 \times 10^{20} \, seconds^{-1}$. This excellent time standard is the frequency of a free electron vibrating at rest. An agreement to make the number exact will define the time unit of the $second$ without referring to years, days, and minutes, or atomic clocks based on much more complicated theory. Indeed the resolution of an atomic clock is based on an approximate theory for the lifetime of certain spectral lines, while the lifetime of an electron appears to be infinite. Perhaps in the future the electron mass - or an easier to control atomic mass - will directly define the standard of time without redundant intermediates.

We have reviewed how arbitrary standardization not coming from Nature affects Newtonian unit conventions. Only one more step of specifying the particular mass convention is needed to {\it derive} Planck's constant.

\subsection{Enter Kilogram} 
\label{sec:enterK}

By choosing a particular convention the conversion factor from our units of $mass$ and $energy$ to engineering units becomes specified. In brief: \begin{itemize} \item Directly from fitting quantum data $\lambda_{e} \sim 3.86 \times 10^{-11} \, cm$. The corresponding frequency scale in the quantum Lagrangian is $\omega_{e} =c/\lambda_{e} \sim 7.8 \times 10^{20} s^{-1}$.  \item The Newtonian electron mass parameter $m_{Ne} = \omega_{e}/c^{2}= 0.865 \,  s/cm^{2}$. This is a convenient macroscopic value. \item By ratios, the proton and Hydrogen mass is $m_{H} \sim 1836 \, m_{e} \sim 1588 \, s/cm^{2}$. \item To a good approximation, one mole ($6.02 \times 10^{23}$) of protons defines a {\bf{gram}} of mass, to relative errors of a few parts per thousand. The arbitrary human conventions of the ``kilogram'' enters {\bf{here}}. It is also possible to define the kilogram by a reference standard object (``artifact''), and deduce Avogadro's number from the mass of a mole. 
\im In $cgs$ convention 1 $erg =1 \,  gm \, cm^{2}/s^{2}$, or  \ba 1\, erg &=&  1600  { s \over cm^{2} \, proton} 6.02 \times 10^{23} { protons \, cm^{2} \over s^{2}} \nn \\ &\sim &   9.63 \times 10^{26} \, s^{-1}. \nn \ea The inverse relation is \ba  1\, {rad \over s} &=& 1.05 \times 10^{-27} \, erg. \label{hbar} \nn \ea The last line defines the conversion constant $h$ going from frequency to $MKS$ units. \eit 

We have not bothered with high precision in the calculation. Unit conventions should be expressed with exact numerical values standardized in their definitions.

\subsection{Paradoxes, Measurements, Group Generators, Gravity}

\label{sec:measure}

Quantum theory is a large subject, so that challenging any one point can lead to distinct types of disagreement about what quantum theory actually implies. We granted early that anyone wanting to keep $\hbar$ in their conventions can do so without contradiction, although we find it redundant. 

What are more troublesome are false paradoxes and sometimes obstacles thrown up to maintain a pedagogical tradition. We are not particularly concerned with paradoxes affecting the ``old'' theory's intricate need to maintain Planck's constant, since as we mentioned we have a new theory. We plan to give a more comprehensive review of those fascinating side issues elsewhere\cite{inprep}, restricting our discussion here.

The basic algorithm to explore questions goes as follows: Every traditional quantum mechanical formula involving $\hbar$ is either an independent fact of mathematics, or our theory, which has been multiplied by some power of $\hbar$ on both sides. 

Most physicists agree that equations are unchanged in content by multiplying both sides by the same constant. Yet a curious degree of resistance is sometimes found to applying the algorithm after its effects are realized, almost as if $\hbar$ should get a special variance from the rules of algebra. We will highlight a few topics briefly to contrast the absence of Planck's constant in {\it our particular} quantum theory: 
\begin{itemize}  

\item The historical path to quantum mechanics invariably begins with Planck. Reviewing his original paper, 
it is interesting that Planck himself could not find his constant from his own analysis of black body spectral data. What Planck's paper literally found\cite{planckoriginal} was a ratio $h/k_{B}=4.866 \,  .\, 10^{-11} sec \,degree$. At that moment Planck was in a position to eliminate both constants in terms of a frequency parameter $\omega_{T}$ for the temperature: a spectrum involving $(exp(\omega/\omega_{T})-1)^{-1}$. However Planck's archaic definitions of energy from thermodynamics made it impossible. His deep commitment to (and invention of) Boltzmann's constant\footnote{Planck's introduction of $k_{B}$ was based on his {\it hypothesis of quanta}, namely molecules. Boltzmann had the gas constant $R$, which needed the hypothesis of quanta and a value for Avogadro's number to reach $k_{B}$. Planck\cite{PlanckBook} introduced the formula for entropy $s =-k_{B} \, log(W)$ as the true ``thermodynamic'' probability, sharply distinguishing it from Boltzmann's $s =-log(W)$ as a mere ``mathematical probability.'' } made it impossible for him to see either constant as avoidable.

\item The historical path often cites Einstein's approach to the photoelectric effect. The``production of cathode rays by illumination of solids'' was but one topic in a phenomenological paper citing several reasons for a new picture of particle-like quanta. Einstein found that Planck's conversion factor of energy and frequency was consistent with a 1902 experimental paper of Lenard. Einstein wrote\cite{photoEin} ``To see now whether the relation derived here agrees, as to order
of magnitude, with experiments, we put $P'=0$, $\nu=1.03 \times 10^{15}$, (corresponding to the ultraviolet limit of the solar spectrum) and
$\beta =4\cdot 866 \times 10^{-11}$. We obtain $\Pi \times 10^{7} = 4.3 \,Volt$, a result which agrees, as to order of magnitude, with Mr. Lenard's results.'' The paper uses Planck's constant $R\beta/N$ as Planck did, with $R$ the gas constant and $N$ the number of ``real molecules'' in gram-equivalent units (per mole.) Symbol $P'=0$ sets the zero of the work function, and $\Pi $ is the voltage to reduce the photocurrent to zero, i.e. the photo-electron's energy.

Just as with Compton's experiment, Einstein's use of conservation of frequency expressed the kinematic fact of time-translational invariance in photoelectric scattering\cite{lamb}. However at this early point the conventional unit of the $Volt$ had not been converted to frequency. We convert without the intermediary of $R$, $N$, $\beta$ or $h$ as follows: From Eq.\ref{balmer} the Hydrogen ionization frequency $ \omega_{E}=2.07 \times 10^{16}\ra 13.6 eV$ makes a fiducial {\it definition} of the volt. It allows re-scaling frequency to frequency, from which Einstein's 4.3 $Volt= (4.3/13.6) \times 2.07 \times 10^{16}/s =6.55 \times 10^{15}/s$. Compare $2 \pi \nu =6.47 \times 10^{15}/s$; the numbers agree within about 1\%. The upshot is the photoelectric effect does not need Planck's constant.

\item Reference to Planck's constant involving ``measurement theory'' are common. A finite value of $\hbar$ is cited as responsible for lower bounds on disturbances caused in measurement. The example known as Heisenberg's microscope it typical. The key steps of post-quantum measurement are different, and involve projecting onto a wave function $|\psi_{2}>$ given a wave function $|\psi_{1}>$. There is a convenient identity \ba |\psi_{1}>=|\psi_{2}><\psi_{2}|\psi_{1}>+ |\psi_{\perp}>,\nn \ea where $|\psi_{\perp}>$ is in the orthogonal complement to $|\psi_{2}>$. Since $|\psi_{2}><\psi_{2}|$ is a normalized projector, the equation says that $<\psi_{2}|\psi_{1}>$ is the pre-existing amount of $|\psi_{2}>$ already presenting in $|\psi_{1}>$. The squared overlap $|<\psi_{2}|\psi_{1}>|^{2}$ is identified as the probability of $|\psi_{2}>$ given $|\psi_{1}>$. When this describes an experiment, the measurement had to be sufficiently gentle that the pre-existing projection is found {\it without} mixing in $|\psi_{\perp}>$ or any other disturbing effects. In other words, any ``uncontrollable disturbance'' of measurement will {\it not} be so simple that mere projection describes it. Whether one uses wave functions or density matrices for observables, Planck's constant does not appear in the Born rule. 

\item In certain cases the fixed normalization of wave functions $<\psi|\psi>=1$, coupled to an absolute value of Planck's constant, is thought to be responsible for ``quantization of energy levels''. It is true that boundary conditions enter quantization but not the overall scale. If the quantum wave of an infinite square well, say, were interpreted classically, the overall amplitude would enter the total classical energy. Once the wave function is normalized the energy is fixed. Here again we must recognize {\it quantum homogeneity symmetry}, which tells us the square well frequencies are the energies no matter how the normalization is set. This is admittedly a radical revision of the concept of energy, but not explained (nor improved) by insisting it be expressed in any particular system of units.

\item Questions of the type, ``Without $\hbar$ how are you going to quantize the harmonic oscillator'' refer to schoolbook exercises. They have little to do with Nature. When setting up new models the parameters tend to come from previous models (values of $\omega_{e}$, etc.) and whatever fudge-factors are needed in the lab. The question ``could we find a primed-system where $i\hbar' \dot \psi =\hat H' \psi$ would give us a new value of $\hbar'$'' has been answered repeatedly. Physicists fit Hamiltonians to data using the existing unit conversion factors, by convention. Exact agreement of one universal value of Planck's constant has been made trivial by universal practice: if not universal discussion. 

 \item Textbooks\cite{sakurai} tend to cite the Stern-Gerlach experiment as yielding an inexplicable quantization with directly observable units of $\hbar$. The experiment is truly inexplicable in a straw-man context of Newtonian point particles there is no reason to consider. In the context of the Schroedinger-Pauli equation one computes solutions to scattering off a non-uniform magnetic field. Solutions predict that beams with orthogonal spins separate. Since $\hbar$ is absent in the equation it is not in the solutions. The fact a particular polarization is strictly correlated with each beam is mathematically true and interesting. We cannot find new information in predicting it by an external principle that the eigenstate of the Pauli spin operator $\hbar\sigma_{z}/s$ is``measured.'' It is circular, and the same statement that a eigenstate of $\sigma_{z}/2$ appears in the correlation. 
 
\item Quantization of angular momentum is cited in elementary treatments as a measure of $\hbar$. From Noether's theorem in the Schroedinger model, the operator yielding orbital angular momentum is $\vec L=-i \vec x \times \NN$. It is a fact of mathematics that eigenstates of $L_{z}$ have whole number labels and a whole number of nodes. There is no logic in calling this quantization ``also'' a prediction of a Principle or Axiom of physics: once a math fact is a math fact, making it a principle would be redundant. 

By the previous analysis the conversion constant from intrinsically dimensionless form to $MKS$ units originates in macroscopic physics. This may need reiteration so we will explain. In Newton's world with Newton's units it is a great mystery why quantum angular momentum is a whole number of $1.05 \times 10^{-27} erg\,seconds$. We are challenged to explain why it is not $7.05 \times 10^{-27} erg\, seconds$, or some other number, and is not the absolute number meaningful? Our answer is that the existence of whole numbers was explained by the wave theory and counting the nodes of spherical harmonics. It is fine and wonderful that $L_{z}Y_{\ell m} =mY_{\ell m}$, and $\vec L^{2}Y_{\ell m}=\ell(\ell+1) Y_{\ell m}$. By an intricate process the number ``1'' for each unit was converted to $1.05 \times 10^{-27} erg\, seconds$ when humans introduced the gram, kilogram, and Avogadro's number, following Eq. \ref{hbar}. Then we agree the number is important and necessary for commerce and engineering, which surely need the gram, kilogram, and Avogadro's number. We don't need them in fundamental physics, and prefer to designate 1 unit as 1 unit.


\item {\it Spin} is often misidentified as ``coming from'' analogies with Planck's constant. A quantum system with spin is described by a wave function $\psi_{a}(\vec x, \, t)$, where $a$ is the polarization index of the spin representation. Although representation theory came into physics after Planck's constant, it is absolutely independent and stands on its own. The mathematics of the rotation group predicts $2 \times 2$ generators acting on a spinor space with the algebra \ba [{\sigma_{i} \over 2}, \, 
{\sigma_{j} \over 2} ]= i \epsilon_{ijk}{\sigma_{k} \over 2}. \nn \ea It is an empty act of notation to define $\vec S = \hbar \vec \sigma$. The notation predicts \ba  [ \,S_{i},  \, 
S_{j}  ]= i \hbar \epsilon_{ijk}S_{k}. \label{commuteS} \ea A finite rotation is then expressed by $U(\vec \theta)= exp(i \vec \theta \cdot \vec S/\hbar)$, in which $\hbar$ cancels out. By the same steps, any quantum commutation relation $[\hat A, \, \hat B] =i \hbar \hat C$ is an ordinary algebraic relation $ [\tilde A, \, \tilde B] =i  \tilde C,$ where $\hat A =\hbar \tilde A$, etc. The commutation relations of $\tilde A$ describe a geometry that involves no physical scale.

The elementary cancellation of $\hbar$ in commutation relations is seldom noticed, for reasons that can be explained. In early times the facts of Lie groups and commutation relations were new to physics, and misidentified as ``quantum effects''. The illegal step of taking $\hbar \ra 0$ on the right hand side of Eq. \ref{commuteS} with other symbols fixed was argued to produce the ``classical limit'' including language such as ``in the classical limit operators commute.'' We suggest the language and concepts fail to pass modern quality-control standards, and should be discouraged. There is no consistent sense in which $[{\sigma_{i}  }, \, {\sigma_{j}  } ]=0$. The classical limit is much more subtle than replacing operators by numbers, {\it except for} a brief moment in history when selling quantum mechanics needed it. 

\item The uncertainty principle is a powerful math fact relating the spread of wave 
numbers $\Delta k$ and the spread of size $\Delta x$ of a wave packet: \ba  \Delta 
k\Delta x \gtrsim 1/2. \nn  \ea Multiplying both sides by $\hbar$ gives \ba  \hbar \Delta 
k\Delta x \gtrsim \hbar/2. \label{uncert}  \ea Introducing $\hbar$ made the first time in history where multiplying a math identity by the same constant on both sides was reported to make a new physical principle. It comes from $[ \, x, -i \p/\p x] =i$, which is the trivial identity it appears to be. 

Acknowledging some sarcasm, it seems deceptive in the current millennium to talk about a particle (six canonical degrees of freedom) and simultaneously introduce Eq. \ref{uncert} as a descriptive feature, concealing an infinite number of degrees of freedom that produced the identity. The sensible reason to write $[ \, \hat x, \, \hat p] =i$ as a commutator is to set up a coordinate-free algebra between symbols $\hat x$, $\hat p$. The underlying bracket algebra of a Hamiltonian system is an invariant concept no matter how it is expressed. Expressing Lie group relations with commutators represents high level discoveries of notation, not discoveries about Nature {\it per se}.

\item After the identity $[ \, x, -i \p/\p x] =i$ was discovered useful, it was implemented again by ``field quantization''.
 For every field $\phi(x)$ and its conjugate momentum $\pi(x)$, the equal-time
commutation relations are \ba [ \, \pi(x), \, \phi(x') \, ] = -i \delta(x-x'). \nn \ea This is a sophisticated invariant way of defining symbol $\pi(x)= -i \delta/ \delta \phi(x)$, to wit, an identity in which canonical momenta are generators of canonical coordinates, which is what we mean by momenta. What is new is the promotion of the dynamics to another infinite dimensional space. The proposal is physics, but $\hbar$ is not involved in it. 

On the huge space of quantum fields there are natural time evolution generators $\hat \Omega = \hat \Omega( \pi, \,\phi) =\hat \Omega(-i \delta/ \delta \phi(x), \, \phi(x))$. That is so general it excludes very little. Meanwhile historical models put great faith in local quadratic functions of $\pi$ and $\phi$ which paid off. 
Whether or not local quantum field theories are a good model of the Universe, ``quantization'' does not need Planck's constant in our approach. 

\subsection{Gauge Theories}
\label{sec:gravity}

Relativity and gauge invariance explain a small puzzle in our development. In Section \ref{sec:dynamics} we found the so-called Coulomb interaction function $U= 
\tilde W/(2 \omega_{*} ) \ra \a/r$. To maintain the same number $\a$ for systems of different $\omega*$, as observed, requires our initial symbol $\tilde W$ to be proportional to $\omega*$. There is no obvious motivation for this in a generic non-relativistic effective field theory. 

The relativistic gauge-covariant derivative explains the puzzle. In simplest form replace $(i \p/\p t)^{2} \ra (i \p/\p t - e A^{0})^{2}$.  With $i \p/\p t \ra \omega_{*}$ as first approximation, the expansion to the non-relativistic domain produces $\tilde W \sim  2 eA^{0}\omega_{*}$ just as needed for a universal constant $\a$. In any event, the relativistic field theory rather than basic quantum mechanics becomes the arena to determine what constants are universal. 

\item Every known force is due to a gauge invariance of one kind or other. General relativity predicts that gravity couples the curvature tensor $R_{\mu \nu}$ to the energy momentum tensor $T_{\mu \nu}$: \ba  R_{\mu \nu} -{1 \over 2}g_{\mu \nu}R = -{G_{N}' \over 8 \pi} \, T_{\mu \nu}. \label{ein} \ea 
The dimensions of $R$ are inverse length-squared, and the dimensions of $T_{\mu \nu}$ determine those of $G_{N}'$. Specifying the action $S$ fixes the rest: \ba S &=& {1 \over 16 \pi G_{N}'}\int d^{4}x \, \sqrt{g} R  \nn \\ &+&  \sum_{j} \, m_{j} \,  \int d\tau_{j} \, {\p x^{\mu} \over \p \tau_{j}}g_{\mu \nu}(x_{j}) {\p x^{\nu} \over \p \tau_{j}}. \nn \ea Thus $G_{N}'$ has units of square-meters. Since this scale is non-trivial, there is a good motivation for seeking new physics in it. The units and size of Newton's constant in $\slasha M \slasha K S$ units ($c=1$) is \ba G_{N}'= 2.5 \times 10^{-64} cm^{2} \nn \ea The scale stands on its own, and does not need Planck's constant; It is refreshing to find no reason quantum mechanics needs to be relevant. It is interesting that the root-inverse of Newton's constant in $\slasha M\slasha K S$ units is a macroscopically large number not far from the 14 billion parsec size of the observable Universe: up to a relatively small factor of about 1500 that might be possible to explain. 


The non-relativistic coupling of gravity to matter is well-known. Since we have dispensed with Newtonian mass, it is interesting we do not need the concept of gravitational mass to express the coupling. If there are $N_{1}$ ($N_{2}$) localized quantum systems (particles) of scale $\lambda_{1}$ ($\lambda_{2}$) separated by $r$, the equivalent interaction function is predicted to be $G_{N}'c\, N_{1}N_{2}/(\lambda_{1} \lambda_{2}r)$. Referring to $\lambda_{*}$ parameters makes this well-defined in a quantum context. The approximate proportionality to wave numbers $1/\lambda_{*}$, which in Newtonian physics add linearly when weakly interacting systems are composed, finally explains the origin of additivity of Newtonian mass in the form of ``weight.''
 \end{itemize}

\section{Precision Fundamental Constants}

\label{sec:precision}

\begin{figure}[htbn]
\begin{center}
\includegraphics[width=3.5in]{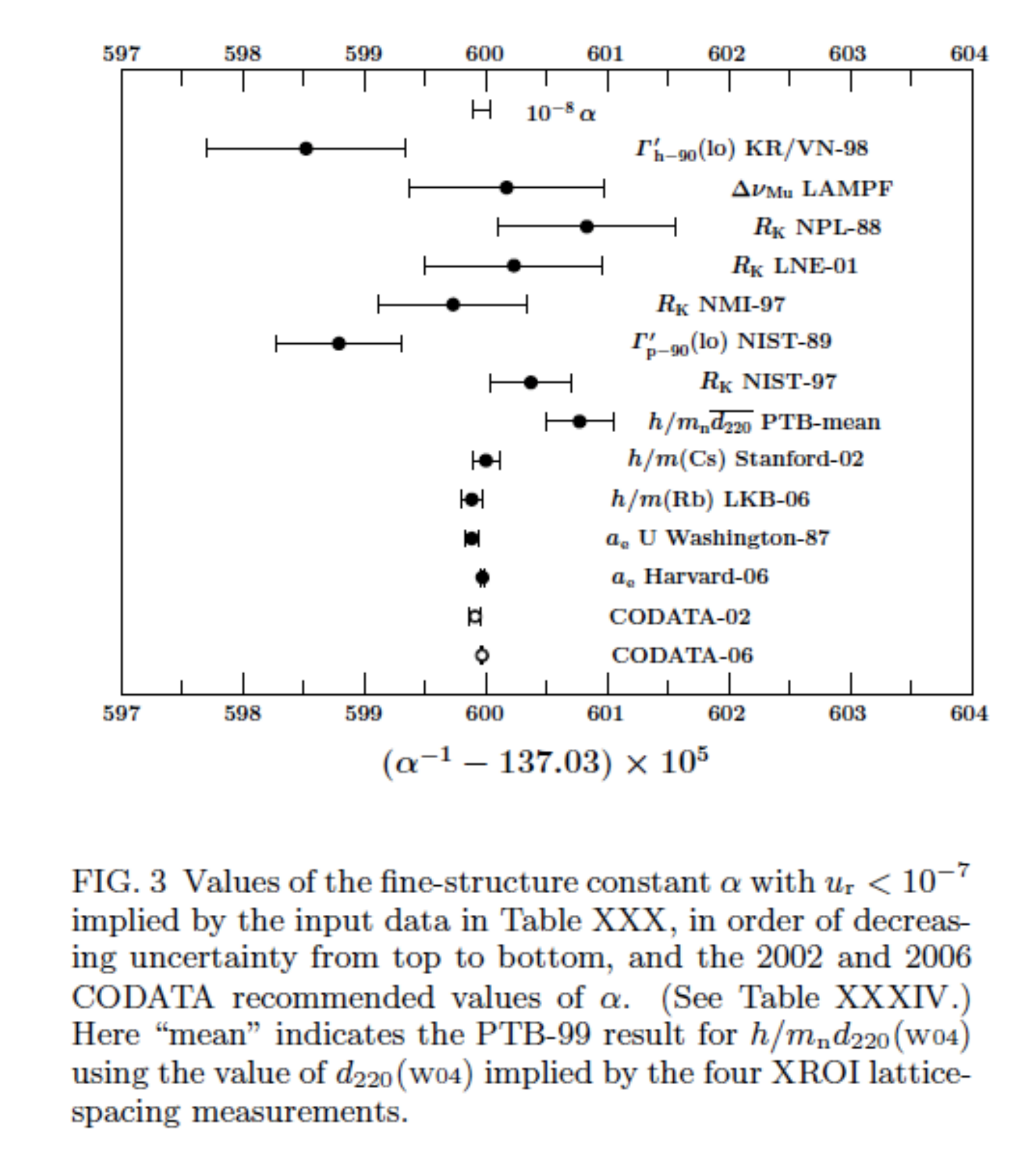}
\caption{ \small Compilation of the values and errors on the fine structure constant $\a$ from Ref. \cite{codata}. Over time $\a$ has become determined better and better by experiments tying its definition to {\it frequency} just as we suggest. }
\label{fig:AlphaData.pdf}
\end{center}
\end{figure}

Great care is mandatory in the field of fundamental constants, but there has always been two separate purposes. One purpose is standardization for commerce and engineering. Another purpose is testing fundamental physics. The two purposes are not the same, because the engineering usage is {\it obliged} to focus on the kilogram and Planck's constant, while fundamental physics (in our opinion!) is obliged to base nothing on them. 

Before continuing we note the exceeding care for consistency that has become standard in perturbative calculations in quantum field theory. In that regime coupling constants are defined in terms of definite subtraction schemes, renormalization points, and order-by-order procedures that are a world of their own. Those issues tend to be buried in the theory-blind standardization of fundamental constants, but they are not trivial. As far as we can tell, our approach has no effect on renormalization conventions, due to an unstated agreement everywhere not to renormalize Planck's constant. 

Consider the electron mass evaluated by the CODATA group\cite{codata}. The uncertainty of the 2010 determination is $4.0 \times 10^{-38}$ kg. The relative uncertainty ($u_{r}$) is $4.4 \times 10^{-8}$. This uncertainty has changed very slowly with time.

What is used for direct measurement of the electron mass? The most accurate determination of the electron mass\cite{farnham} by Farnham {\it et al} and cited by CODATA-2006 is based on the cyclotron frequency of {\it classical electrons} orbiting in Penning traps. We observe that a {\it classical} model of Newtonian electrons is a theory subject to 
numerous assumptions. The dynamics of particles in magnetic fields are not Newtonian, and the characteristic lifetime and frequency spread of cyclotron orbits due to synchrotron radiation is but one of the issues clouding the interpretation of the ``mass'' deduced from data. Farnham {\it et al} state that the Penning traps work best with 5-13 electrons. It is seldom discussed that extending microscopic quantum propagation into the classical regime is exquisitely sensitive to uncontrolled tiny effects. That is because the ``Ehrenfest relations'' so convincing for a beginning treatment of ``free'' Schroedinger particles do not lead to high-precision theory of interacting particles, as far as we know. How to precisely formulate the concept of 5-13 Newtonian electrons does not appear in the references. 

What about Planck's constant? The {\it Overview} of the unpublished CODATA2010 adjustments states that a new value of Avogadro's number obtained from highly enriched silicon has a $u_{r}$ of $3 \times 10^{-8}$, providing an inferred value of $h$ with essentially the same uncertainty. This (indirect) uncertainty is somewhat smaller than $u_{r}$ of $3.6 \times 10^{?8}$ of the most accurate directly-measured watt-balance value of $h$. Yet the two values disagree. That has led to a recommended $u_{r}$ of $h$ of $4.4 \times 10^{?8}$, which is almost no change. This value coincides exactly with the relative error on $m_{e}$ just discussed.  

Since it is a matter for specialists we are in no position to compete with the goals and methods of dedicated groups such as CODATA in standardizing fundamental constants. On the other hand it is valid to estimate the effects on fundamental constants by modestly revising theoretical assumptions. Then, and for fundamental purposes in this Section, we abandon reference to the kilogram, Planck's constant, and the classical electron mass. We define the electron's inertial mass $m_{Ne}  \equiv  hc/2 \pi \lambda_{e}$ to be an identity. The identity allows fixing $h$ to an exact reference value. As reviewed with Eq. \ref{egone} the constant of electric charge $e$ also does not appear in quantum physics, and the definition $e = \sqrt{ \hbar c \a} \ra \sqrt{\a}$ is also taken as an identity.  

It is worth mentioning that our decision involves physics, and some future technology might find the classical electron mass with superb precision, adding information. Either that process would confirm our decision on theory, or contradict it. If contradicted then the meaning and values of fundamental constants will evolve once more. We are concerned here with what can be accomplished within the theory of this paper. 

\subsubsection{Preliminary Constant Values} 

\begin{figure}[htbn]
\begin{center}
\includegraphics[width=4.5in]{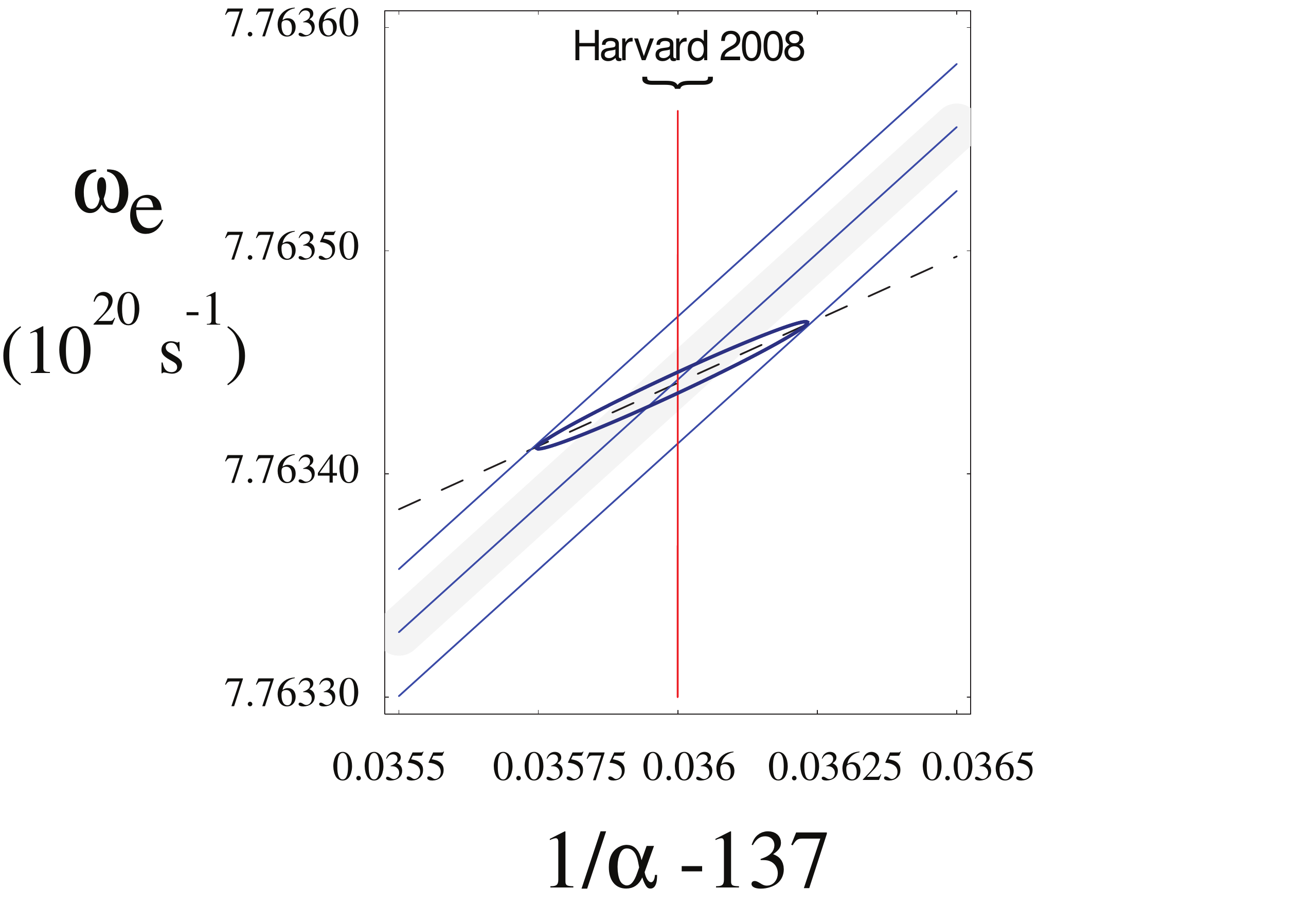}
\caption{ \small  Improved determination of $\omega_{e}=c/\lambda_{e}$ and $\a$ from data. The dashed line shows Eq. \ref{valone} from the Rydberg constant, whose errors are thinner than the line width. Information from the hyperfine interval of positronium with errors is shown by the shaded band\cite{adkins,positronium}. The intersection (region inside ellipse) determines $\a =137.0360  \pm .00025$ and $\omega_{e}=( 7.763 \pm 0.00002 ) \times 10^{20}s^{-1}$. A recent experiment (Harvard 2008) fixes $\a$ with great accuracy (errors thinner than the vertical (red online) line), which then fixed $\omega_{e}$ and the electron mass with errors smaller than any previous determination. }
\label{fig:omegaalpha}
\end{center}
\end{figure}

\begin{table}[htbn] 
  \centering 
\resizebox {.9\textwidth }{!}{ 

 \begin{tabular}{ccccc}
\hline \\
 $Quantity$ &  $  Method$  &  $u_{r}$ &  $u_{R} CODATA_{2006}$ & $u_{R} CODATA_{2010}^{*}$   \\ \\
   \hline \\ 
    $m_{Ne}$ & $R_{\infty}(\a_{2006})$ &  $1.4 \times 10^{-9}$ & $5\times 10^{-8}$ & $4.4\times 10^{-8}$ \\ 
   $m_{Ne}$ & $R_{\infty}(\a_{2008})$ &  $7.4 \times 10^{-10}$ &  $5\times 10^{-8}$   & $4.4\times 10^{-8}$ \\ 
   $e$  &  $\sqrt{\a_{2006}} $ & $3.4 \times 10^{-10}$  & $2.5 \times 10^{-8}$ & $...$ \\
    $e$  &  $ \sqrt{\a_{2008}} $ &  $1.8 \times 10^{-10}$ & ... &  $2.2 \times 10^{-8}$\\ 
    $\hbar$ & $exact$  & $0$ & $5 \times 10^{-8}$ & $4.4 \times 10^{-8}$. \\ \\ \hline
 \end{tabular} }

  \caption{ \small  Summary of relative uncertainties $u_{r}$ in the values of fundamental constants using our procedure.  $^{*}$ CODATA-2010 analysis is unpublished, but available on website.} 
  \label{tab:values}
\end{table}

While determining fundamental constants generally uses global fits to many variables, two independent data points suffice to determine two constants of the theory. The Rydberg constant $R_{\infty}= m_{Ne} c\a^{2}/2h$ and the fine structure constant $\a$ will predict $m_{Ne}$. We choose $R_{\infty}$ because its relative uncertainty is of order $5 \times 10^{-12}$, which is so small it can be neglected. There are a number of ways to get high precision data for $\a$.

Consider the ground state hyperfine interval of positronium\cite{adkins}. Ref.\cite{positronium} reports the experimental frequency shift \ba \Delta f =  2.0338910(74) \times  10^{11} \, s^{-1} . \label{rel1} \ea  Ref.\cite{adkins} cites a calculation in perturbation theory \ba \Delta f &=& {\a ^4  \omega_{e} \over 2 \pi}  \nn \\ & \times & ( \,  7/12 - \a /\pi(1/2 ln(2)+ 8/9 ) \nn \\ &+& 5/24 \a ^2 ln(1/\a ) +0.6 \a ^2  \, ) . \label{shift} \ea Two constants are determined, as shown in Figure \ref{fig:omegaalpha}. The errors in $\Delta f$ dominate, yielding \ba {1\over \a}  & =& 137.0360  \pm .00025;  \nn \\ \omega_{e} & =& ( 7.763 \pm 0.00002 ) \times 10^{20} s^{-1}. \nn \ea This illustrates how the constants can be determined to a relative accuracy of few parts of $10^{-6}$ without great complication.

The electron magnetic moment parameter $g-2$ is arguably more reliable. A 2006 Harvard study\cite{harvard} found \ba 1/\alpha_{2006} = 137.035 999 711(96), \nn \ea a relative uncertainty of $ 7.0 \times 10^{-10}$. It is worth noting that $g-2$ is measured directly in terms of a {\it frequency}, explaining why the existing uncertainties in $m_{e}$ do not degrade this determination of $\a$. When the 2006 errors of $\a$ are applied to the vertical line in Fig. \ref{fig:omegaalpha} the errors are too small to be visible. It leads to a relative error in $m_{Ne} $ of about $1.4 \times 10^{-9}$. Compared to the published CODATA-2006 (unpublished 2010) results (Table \ref{tab:values}), the value of $m_{Ne} $ is determined about 36 (31) times more precisely by our method. 

In 2008 the Harvard group\cite{harvard2008} announced improvement of combined theoretical and experimental uncertainties of $\a$ to 0.37 parts per billion: \ba  {1\over \a}= 137.035 999 084(51).\nn \ea Adopting this figure produces the electron mass \ba m_e = 9.10938 215 00(70) \times 10^{-31} \, kg. \nn \ea The relative uncertainty of $7.4 \times 10^{-10}$ is 67 times less than reported by the best previous value published by CODATA-2006. 

Our uncertainties of the electric charge $e$ are simply determined by $\Delta e/e \sim (1/2)\Delta \a /\a$. Using the 2008 value of $\a$, the standardized values of $c$, $\hbar$ and the electrical constant $\epsilon_{0}$ of $MKS$ units we find  \ba e = 1.60217 648 684(26) \times 10^{-19}. \nn \ea The relative uncertainty $u_{r}=1.8 \times 10^{-10}$. The numerical value is within the error bars of the CODATA-2010 determination that cites a relative uncertainty of $2.5 \times 10^{-8}$. Compared to the published determination of $e$, our relative uncertainty is $2.5 \times 10^{-8}/1.8 \times 10^{-10} \sim 139$ times smaller: see Table \ref{tab:values}.

We have shown that the electron mass and electric charge are substantially improved when reliance on the kilogram and conversion factors of Planck's constant are avoided. This is related to the generally known fact that the ratio of the electron mass to an atomic mass unit (Carbon-12) can be determined with greatly improved uncertainty by using a common method (Penning trap) and avoiding the kilogram. It is no accident that what is actually measured in these experiments are comparisons of {\it frequency} $\omega_{e}$ versus {\it frequency} $\omega_{^{12}C}$, just as we find is fundamental.  

\subsubsection{Correlated Parameters} 

What about discrepancies between calculations in perturbative $QED$ and data, including the anomalous moment measurements we have just used for our preliminary constant determinations? They are welcome! One of the main interests in precision fundamental constants is in testing new physics. Clinging to outmoded theoretical procedures happens to introduce inconsistencies. If a disagreement between theory and experiment is sharpened by complete consistency of definitions it can only represent progress.


One might ask whether our determination lost information compared to finding $\hbar, \, e, \, m_{e}, \, \a$ separately.  It might seem that separate determination and comparison would be ``testing the theory''.  However the process of testing theories means posing alternative physical hypotheses capable in principle of giving different answers. Groups such as CODATA are not charged with considering models of new physics. The default physics has become one uniform framework of quantum electrodynamics, with theoretical contributions from other sources, which is the same framework as ours. Rather than global fits testing a framework, global fits will produce global error bars. 

Yet sometimes the data fitting process can indicate redundant parameters. That is done by finding stalemates with unusually high parameter degeneracy. A careful reading of the 105 page CODATA2006 document\cite{codata} will find the experts are aware of a strong correlation between the quantities we have identified. Great effort has gone into studying the correlations between the {\it experimental} inputs, which are seldom independent. We are concerned with the outputs. Table $LI$, page 102 of Ref. \cite{codata} shows the correlation of the evaluated $h$ and $m_{e}$ is $r  =  0.9996.$ (The correlation goes to $r=0.9999$ in the 2010 website material.) While it may be simplistic, this correlation is an outcome predicted by the symmetry of Eq. \ref{symmy}, varying $m_{e}$ and $h$ while keeping the observable Rydberg frequency $R_{\infty}= m_{Ne} c\a^{2}/2h$ fixed: Exactly as we did in Section \ref{sec:fitting}. The other correlations in Table $LI$ with magnitudes exceeding $0.999$ come between $(h,e)$, $(h, N_{A})$, $(N_{A}, m_{e})$, and $(N_{A},e)$, where $N_{A}$ is Avogadro's number. These facts support our view that eliminating the correlations is actually a matter of principle, not improving technology. 
 
Trimming redundant constants such as $e$ and $\hbar$ clears the way for meaningful constants to be determined better, and move on to compare precision experiments better. The need for many independent experiments remains. For example, the theory of the Josephson effect has traditionally been formulated in terms of a constant $K_{J}=e/h$. This can be traced to theoretical decisions to separate the parameter $\a$ found in quantum electrodynamics into terms involving $e$, and an external magnetic field also proportional to $e$, representing a net dependence on $e^{2}/h$. Consistent definitions should improve tests of the theory of Josephson junctions and quantum Hall physics. The unsettled discrepancy between the electron and muon magnetic moments is another example where high precision constants are important.
 
\subsection{Exit Kilogram}

\begin{figure}[htbn]
\begin{center}
\includegraphics[width=3.5in]{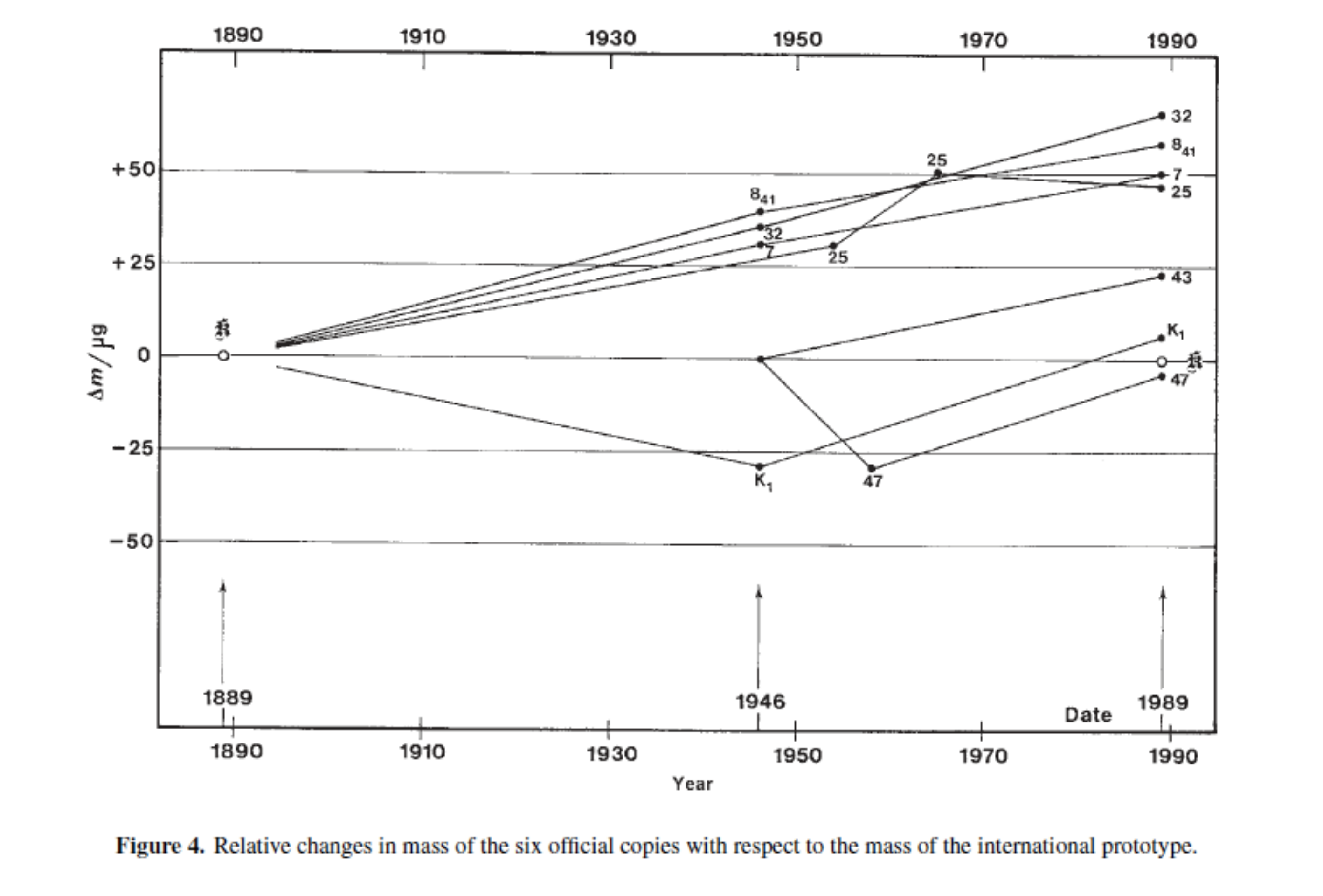}
\caption{ \small Observed variation of mass with time for the International Prototype Kilogram ($IPK$) relative to 6 official copies.  Scale ``0'' is relative to the $IPK$, which has not so far changed relative to itself. Planck's constant is changing with time by an amount linear in the change of the standard masses. From Ref. \cite{massdef}. }
\label{fig:masstime}
\end{center}
\end{figure}

\begin{figure}[htbn]
\begin{center}
\includegraphics[width=3.5in]{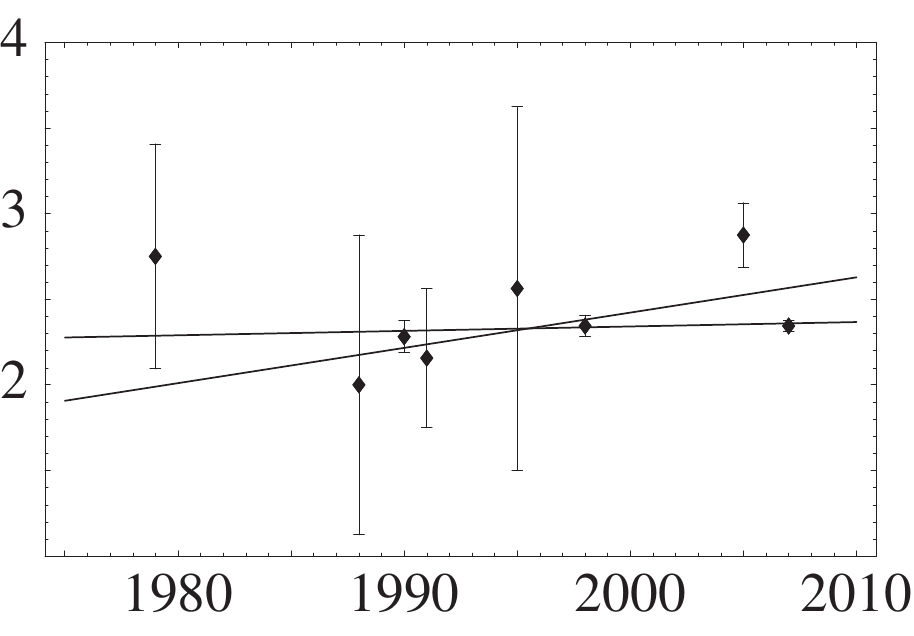}
\caption{ \small Time variation observed in Planck's constant. Vertical scale shows values of $10^{5}(h/10^{-34}-6.620)$ from Ref. \cite{codata}. The fit with the smaller (but non-zero) slope is close to the slope predicted by time variation of the International Prototype Kilogram, Fig. \ref{fig:masstime}. The fit with the larger slope omits the most recent point, whose tiny errors make the global fit worse. }
\label{fig:PlanckTime.eps}
\end{center}
\end{figure}

Finally we return to the kilogram. The kilogram is the only $SI$ unit not defined by independently reproducible experiments. In comparison with 6 official copies, the single International Prototype Kilogram has consistently lost mass over time\cite{massdef}. An unknown mechanism is causing a loss rate $|\Delta m/m| \sim 2.5 \times 10^{-7}/century$ (Fig. \ref{fig:masstime}). The relative change of the mass standard itself every ten years is comparable to the relative errors on the electron mass found in the last ten years. 

Time-dependence of fundamental constants has been an important topic since Dirac\cite{timedep} suggested the fine structure constant might be time dependent. Fig. \ref{fig:PlanckTime.eps} shows a fit to the time dependence of $h$. The slope of the minimum $\chi^{2}$ fit is $\Delta h/h =3.3 \times 10^{-7}/century$. It is remarkably close to the slope of the $IPK$. This is perhaps fortuitous. The most recent point with the smallest errors dominates the fit. If this point is removed, the value of $\chi^{2}$ decreases by 4 units, indicating a much better fit from removing an outlier. The slope of Planck's constant {\it increases} to $\Delta h/h =3.0 \times 10^{-6}/century$. Planck's is the first fundamental constant to develop an observed time-dependence. 

While interesting we find it foolish to take this seriously. But it is supposed to be foolish to maintain forever that $\hbar$ must be fundamental now because it was once fundamental in the past. We find it simple and rational to fix Planck's constant to a definition. It fulfills the long-standing goal of a reproducible standard of mass\cite{planckdef} not intrinsically depending on comparison with artifacts. 
  
\section{Concluding Remarks}
 
Planck's constant entered physics by a particular historical path. Newtonian concepts of energy and inertial mass were assumed. To this day the historical path is used in teaching the subject. In the meantime fundamental physics has evolved. Research practice has every reason to drop holdovers from history. The introduction suggested that the agreement to drop Planck's constant from fundamental units should not be difficult. Some physicists accustomed to ignoring it might not find the conclusion trivial, yet that appreciation is hardly universal. We have shown that unless Newtonian physics has primacy and new information not available from quantum theory, reference to Planck's constant is redundant. When historical prejudices are dropped Planck's constant disappears. 

There is still a place for standardizing Planck's constant, just as standardizing other units is important to engineering and commerce. Standardization of $\hbar$ and $c$ share the common element of removing barriers to precision measurements of other constants. Unlike $c$, which in principle might disagree with theory, Planck's constant in our quantum theory is unobservable, and we can't even suggest an experiment to find it. The challenge of testing fundamental physics should not be saddled with constructs set up by human conventions. It would be delightful if the tradition of retaining Planck's constant might not forever propagate into tests of fundamental physics. Then the current generation of precision quantum measurements might find discrepancies in the fundamental quantum parameters requiring new physics. \\ 

{\it Acknowledgements:} Research supported in part under DOE Grant Number DE-FG02-04ER14308. We thank Carl Bender, Don Colloday, Jacob Hermann, Danny Marfatia, Phil Mannheim, Doug McKay, Dan Neusenschwander, and Peter Rolnick for comments.

\end{document}